%% file: document.tex
\documentclass[journal,10pt,twocolumn]{IEEEtran}

\usepackage{booktabs}
\usepackage{multirow}
\usepackage{comment}
\usepackage{algorithm}
\usepackage{algorithmic}
\usepackage{graphicx}
\usepackage{hyperref}
\usepackage{url}
\usepackage{lscape}
\usepackage{booktabs}
\usepackage{multirow}
\usepackage[caption=false]{subfig}
\usepackage{diagbox}
\usepackage{listings}
\usepackage{diagbox,colortbl}
\usepackage[overload]{empheq}
\usepackage{amsmath,amsfonts,amssymb,mathtools}
\usepackage{xr}
\usepackage{balance}
\usepackage[backend=bibtex,citestyle=numeric-comp,bibstyle=ieee,sorting=none,minbibnames=1,maxbibnames=3,defernumbers=true,firstinits=true]{biblatex}

\definecolor{mygray}{gray}{0.5}

\title{Dictionary Attacks on Speaker Verification}

\author{
  Mirko Marras~\IEEEmembership{Member,~IEEE}, Paweł Korus~\IEEEmembership{Member,~IEEE}, Anubhav Jain and Nasir Memon,~\IEEEmembership{Fellow,~IEEE}
  \thanks{This material is based upon work supported by the National Science Foundation under Grant No. 1956200.}
  \thanks{M. Marras is with the Department of Mathematics and Computer Science, University of Cagliari, Italy (e-mail: mirko.marras@acm.org).}
  \thanks{P. Korus is with Amazon. The work was done while at Tandon School of Engineering, New York University, USA, and also the Department of Telecommunications, AGH University of Science and Technology, Poland (e-mail: pkorus@agh.edu.pl).}
  \thanks{A. Jain and N. Memon are with Tandon School of Engineering, New York University, USA (e-mail: aj3281@nyu.edu; memon@nyu.edu).}
}

\makeatletter
\g@addto@macro{\UrlBreaks}{\UrlOrds}
\makeatother

\begin{document}

\maketitle

\begin{abstract}
In this paper, we propose dictionary attacks against speaker verification - a novel attack vector that aims to match a large fraction of speaker population by chance. We introduce a generic formulation of the attack that can be used with various speech representations and threat models. The attacker uses adversarial optimization to maximize raw similarity of speaker embeddings between a seed speech sample and a proxy population. The resulting master voice successfully matches a non-trivial fraction of people in an unknown population. Adversarial waveforms obtained with our approach can match on average 69\% of females and 38\% of males enrolled in the target system at a strict decision threshold calibrated to yield false alarm rate of 1\%. 
By using the attack with a black-box voice cloning system, we obtain master voices that are effective in the most challenging conditions and transferable between speaker encoders. We also show that, combined with multiple attempts, this attack opens even more to serious issues on the security of these systems.
\end{abstract}

\section{Introduction}

Biometric technologies constitute one of the most popular solutions to user authentication. They can offer high reliability and better user experience than classic password-based systems, especially on mobile devices~\cite{DBLP:journals/corr/abs-2105-06625}. Among the plethora of available modalities, the most commonly deployed verification systems look at faces~\cite{DBLP:journals/ijon/WangD21a}, fingerprints~\cite{DBLP:books/sp/MaltoniMJP09}, and speech~\cite{DBLP:journals/spm/HansenH15} - all of which can be used in modern smartphones. In this study, we focus on speaker verification, a key component of voice assistants, which represent a rapidly growing human-computer interaction method popularized by smart speakers~\cite{DBLP:journals/tochi/AmmariKTB19,hoy2018alexa}. 

Like other biometric modalities, speech remains susceptible to attacks~\cite{DBLP:journals/corr/abs-2105-06625} which target both speech recognition (e.g, by crafting hidden voice commands~\cite{DBLP:conf/uss/ChenYZ0Z0020}) and speaker verification (e.g., impersonation via spoofing, re-play or voice synthesis/conversion~\cite{DBLP:conf/icassp/WuKDYSTK15,DBLP:conf/icassp/LiZWYLM20}). Speaker impersonation studied to date exclusively focuses on \emph{targeted attacks}, which make two critical assumptions: (i) there is a specific single \emph{victim} (i.e., a target identity whose voice the attacker tries to imitate) and (ii) a sample of the victim's voice is available (or needs to be obtained). While the required sample size varies, and tends to change depending on the attack method and authentication protocol (e.g., text-independent~\cite{DBLP:journals/ejasp/BimbotBFGMMMOPR04,DBLP:conf/interspeech/YouGDD19a,DBLP:journals/jstsp/YamagishiKELT17} or interactive challenge-response~\cite{heigold2016end,variani2014deep}), the principle remains the same.

In this paper, we propose a novel attack vector against speaker verification systems: \emph{untargeted dictionary attacks}. In contrast to targeted attacks, the goal is to match a non-trivial fraction of the user population by pure chance, without any knowledge of the victim's identity or voice. Such an attack could be leveraged for unlocking a phone found on the street or facilitating mass-scale voice commands to voice assistants in compromised home networks~\cite{DBLP:conf/dsn/HeRA21}. Our approach involves adversarial optimization of a novel attack objective and can be applied to arbitrary speech representations (e.g., waveforms, spectrograms, speaker embeddings), making it adaptable to different systems and verification protocols (e.g., text-dependent or independent). This attack opens up a novel threat against the voice modality.

The feasibility of dictionary attacks has recently been shown for the fingerprint~\cite{DBLP:journals/tifs/RoyMR17,DBLP:conf/btas/BontragerRTMR18} and the face~\cite{DBLP:conf/icb/NguyenYEM20} modalities. The inspiration comes from \emph{biometric menagerie}~\cite{DBLP:journals/pami/YagerD10}, a well-established principle of numerous biometrics to exhibit large variations of matching propensity across individuals. In particular, the most relevant group for our work is represented by people who tend to match others easily (\emph{wolves}) and people highly susceptible to be matched (\emph{lambs}). Dictionary attacks aim to exploit this phenomenon to generate \emph{master biometric examples} that maximize the impersonation capability of generated samples. Combined with rapidly improving generative machine-learning models, e.g., generative adversarial networks~\cite{DBLP:conf/nips/GoodfellowPMXWOCB14} or variational auto-encoders~\cite{DBLP:journals/corr/KingmaW13}, this attack may soon create the perfect storm for biometric authentication.

Our study makes the first step to formalize and extensively evaluate dictionary attacks against speaker verification systems. The main contributions of our work are listed below.
\begin{enumerate}
   \item We propose a generic formulation of the attack based on adversarial optimization driven by raw similarity of speaker embeddings. The attack can be applied to various speech representation domains and threat models.
   \item We evaluate the attack, comparing three speech representations and several speaker encoders, under white- and black-box settings, showing strong generalization to an unseen speaker population and (in some settings) non-trivial transferability to unseen encoders.
   \item We show that speaker verification systems are susceptible to this attack and that the effect varies across genders. In our experiments, an accidental intrinsic bias of speaker encoders made female speakers remarkably more vulnerable to the attack.
\end{enumerate}

Compared to our prior study \cite{DBLP:conf/interspeech/MarrasKMF19}, we have revised and generalized the attack to enable seamless application to various speech representation domains. We also extended the evaluation to include several speaker encoders and various threat models. Our version in this paper leads to substantially better results and can be even used in challenging conditions, e.g., to evolve transferable master voices based on black-box access to a third-party voice cloning system with variable output. 

\section{Related Work}

\subsection{Speaker Modelling} 

Speaker recognition involves two main tasks~\cite{DBLP:journals/spm/HansenH15}: \emph{identification} aims to identify the speaker among a set of possible hypotheses; \emph{verification} aims to confirm the identity of the claimed speaker and operates in an open-set regime based on a gallery of enrolled speech samples. Speaker modeling has recently been dominated by deep neural networks~\cite{DBLP:journals/nn/BaiZ21} (DNNs) which remarkably outperform classic solutions like GMM-UBM \cite{DBLP:journals/dsp/ReynoldsQD00} or i-vector~\cite{DBLP:journals/taslp/DehakKDDO11}. DNNs are typically pre-trained for the identification task, but are then adapted to open-set verification by discarding the classification head and extracting a compact intermediate representation, referred to as a \emph{speaker embedding}. The embeddings are then compared between the query and enrolled samples to confirm the speaker's identity.

Speaker enrollment typically involves the collection of multiple speech samples, whose embeddings need to be combined. Some of the traditional methods (e.g., a PLDA model \cite{DBLP:journals/spm/HansenH15}) assume statistical independence, which is hard to achieve in practice. As a result, simpler scoring strategies are often preferred, e.g., averaging the embeddings or taking the one with maximum similarity. A recent study~\cite{DBLP:journals/dsp/RajanAHK14} showed that the average embedding often leads to superior performance, which makes it a popular choice~\cite{DBLP:conf/asru/ZhangK17,DBLP:conf/nodalida/KasevaKRK21}.

Countless model architectures have been proposed for speaker encoding. Some of the most prominent differences involve selection of the input acoustic representation, backbone network, and temporal pooling strategy. While directly using waveforms to learn a representation is possible~\cite{DBLP:conf/slt/RavanelliB18}, it is much more common to use a hand-crafted 2D representation (e.g., spectrograms or filterbanks). The latter enables adaptation of successful backbones from computer vision, e.g., VGG~\cite{DBLP:journals/csl/NagraniCXZ20} or residual networks (ResNet) \cite{DBLP:conf/cvpr/HeZRS16,DBLP:conf/icassp/WangYLF20,DBLP:conf/icassp/YadavR20}. Dealing with the time dimension can rely on recurrence~\cite{DBLP:conf/icassp/WanWPL18}, pooling~\cite{DBLP:conf/icassp/YadavR20,DBLP:conf/interspeech/ZhuKSMP18} or specialized architectural designs. As an example, Time Delay Neural Networks (TDNNs) use a 1D convolution structure along the temporal axis and are adopted in the popular x-vector architecture \cite{DBLP:conf/interspeech/SnyderGPK17,DBLP:conf/icassp/SnyderGSMPK19}. 

Usually, trainable pooling layers achieve better results than simple pooling operators, (e.g., average pooling \cite{DBLP:conf/icassp/YadavR20} or statistical pooling \cite{DBLP:conf/icassp/SnyderGSPK18}). Some of the most successful learned designs include the family of VLAD models. NetVLAD \cite{DBLP:conf/icassp/XieNCZ19} assigns each frame-level descriptor to a cluster and computes residuals to encode the output features. Its variant GhostVLAD \cite{DBLP:conf/icassp/XieNCZ19} improved performance by excluding some of the original NetVLAD clusters from the final concatenation, such that undesirable speech sections are down-weighted.

\subsection{Adversarial Attacks in Speech Processing}
Originally introduced in computer vision \cite{DBLP:conf/nips/GoodfellowPMXWOCB14}, adversarial attacks refer to genuine samples imperceptibly modified by tiny perturbations to fool classifiers with high chance. In the context of speech, this type of attack can be broadly categorized based on the targeted task, i.e., speech or speaker recognition. 
In the former, the goal is to embed carefully crafted perturbations to yield automatic transcription of a specific malicious phrase. In \cite{DBLP:conf/uss/CarliniMVZSSWZ16}, the attacker uses inverse feature extraction to generate obfuscated audio played over-the-air, which allows for issuing hidden commands to voice assistants.
Later, \cite{DBLP:conf/sp/Carlini018} proposed a white-box attack based on gradient optimization, leading to quasi-perceptible adversarial perturbations, finally improved using psychoacoustic modeling \cite{DBLP:conf/icml/QinCCGR19}. To avoid repeated optimization hindering real-time use, a recent work by  \cite{DBLP:conf/interspeech/NeekharaHPDMK19} designed an algorithm to find a single universal perturbation, that can be added to any speech waveform to cause an error in transcription with high probability. Finally, \cite{DBLP:conf/uss/ChenYZ0Z0020} showed that adversarial commands can be also hidden in music. The authors used a surrogate model to create transferable adversarial examples that can achieve this goal.  

Attacking speaker verification systems initially relied on spoofing and replay attacks. Susceptibility to adversarial examples has gained attention only recently. The goal is to craft an attack sample from a voice uttered by a seed speaker, so that it is misclassified as a different one (either specific or any), while still being recognized as the seed speaker by human listeners. In a white-box setting, the FGSM attack made it possible to generate adversarial examples with high success rate~\cite{DBLP:conf/icassp/LiZWYLM20}. \cite{DBLP:conf/interspeech/WangGX20} constrained the perturbation based on a psychoacoustic masking threshold to obtain imperceptible samples. 
To obtain robustness against reverberation and noise,~\cite{DBLP:journals/vlsisp/XieLSLCY21} proposed a gradient-based optimization that generates robust universal adversarial examples (though the attack was not tested over-the-air). \cite{DBLP:conf/sp/ChenCFDZSL21} used a gradient estimation algorithm (NES) in a black-box setting. While the study used a small dataset, the attack had a high success rate in a practical setting.

All of the existing attacks (including both spoofed and adversarial samples) are targeted, i.e., they aim to pass authentication as a specific individual. However, biometric systems exhibit large variations in matching propensity across individuals, which can be exploited to open a novel threat vector. Hence, the untargeted nature of the proposed dictionary attacks is fundamentally different from the untargeted nature of adversarial attacks on machine learning models. In this context, the latter would aim to prevent authentication as a particular person without specifying the desired target identity. 

\subsection{Dictionary Attacks in Biometrics}
Dictionary attacks use prior knowledge about the expected success rate to triage brute-force authentication attempts. They naturally apply to passwords, but until recently have not been considered for other authentication modalities. In biometrics, such attacks are qualitatively different from spoofing and do not require any knowledge about the victim (e.g., speech samples)~\cite{DBLP:journals/speech/WuEKYAL15}. This threat is enabled by large variation in matching propensity across individuals (biometric menagerie~\cite{DBLP:journals/pami/YagerD10}) and further exacerbated by the usability-security trade-offs in mass deployments (e.g., partial finger impressions~\cite{DBLP:journals/tifs/RoyMR17}).

The concept of dictionary attacks in biometrics was introduced only recently. The vulnerability was first demonstrated on fingerprints~\cite{DBLP:journals/tifs/RoyMR17} and subsequently extended to faces~\cite{DBLP:conf/icb/NguyenYEM20}. Initially, an existing fingerprint with the highest impostor score was selected as a \emph{master print}~\cite{DBLP:journals/tifs/RoyMR17}. In the next iteration, synthetic master prints were created by first-order hill-climbing, initialized on the most promising real fingerprints from the first approach. However, local search algorithms may get stuck in local minima or take a long time to converge. \cite{DBLP:conf/btas/BontragerRTMR18} used diversity-quality evolution to address this issue and a generative adversarial network (GAN) to parametrize the search space. The same approach was successful for faces~\cite{DBLP:conf/icb/NguyenYEM20}. 

So far, dictionary attacks have not been studied for speech. Our preliminary work~\cite{DBLP:conf/interspeech/MarrasKMF19} demonstrated that adversarial optimization of spectrograms in a white-box setting consistently increases impersonation rates in VGGVox~\cite{DBLP:journals/csl/NagraniCXZ20}. The resulting adversarial samples could match, on average, 20\% (10\%) of female (male) speakers in an unseen population. In this paper, we generalize our attack and test it against multiple systems and diverse speech representations. We achieved substantially improved impersonation rates and demonstrate non-trivial transferability across speaker encoders. 

\section{Problem Statement and Attack Methods}

\begin{figure*}[t]
   \includegraphics[width=\textwidth]{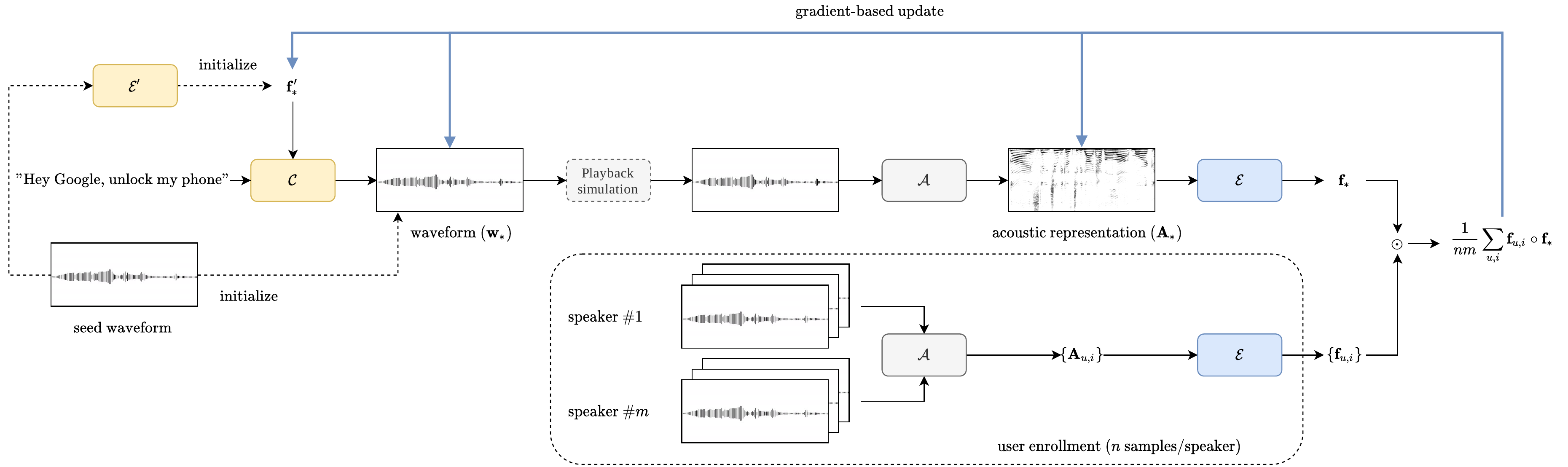}
   \caption{The proposed adversarial optimization protocol for finding master voice samples: (1) a seed sample is used to initialize the attack; (2) a speaker embedding is computed and compared with a collection of enrolled embeddings from a proxy optimization population; (3) the gradient of the adopted similarity metric is computed and leveraged to update a chosen speech representation (e.g., waveform, spectrogram, or speaker embedding); depending on the adopted threat model either full gradient (white-box setting) or its approximation (black-box) is used.}
   \label{fig:optimization}
\end{figure*}

\begin{table}[!b]
   \centering
   \caption{Key symbols and notation summary}
   \label{tab:notation}
   \begin{tabular}{lll}
      \toprule
      waveform & $\mathbf{w}$ & $\in [-1,1]^*$ \\
      acoustic representation & $\mathbf{A}$ & $\mathcal{A}(\mathbf{w}) \in \mathbb{R}^{k \times *}$ \\
      speaker embedding & $\mathbf{f}$ & $\mathcal{E}(\mathbf{A}) \in \mathbb{R}^{e}$ \\
      speech parametrization/generation & $\mathcal{G}(\mathbf{w},\mathbf{v}|\theta)$ & \\
      \midrule
      collection of speaker embeddings & $F$ & $= \lbrace \mathbf{f}_{\mathtt{u},\mathtt{i}} \rbrace \in \mathbb{R}^{n \times m \times e}$  \\
      ... for user $u$ & $F_{u}$ & $= \lbrace \mathbf{f}_{\mathtt{u},\mathtt{i}} : \mathtt{u} = u \rbrace$ \\
      ... for population $U_t$ & $F^{o}$ & $= \cup_{u \in U_o} F_u $ \\
      optimization (proxy) population & $U_o$ &  \\
      test population & $U_t$ &  \\
      \midrule
      \# enrolled users & $m$ & $|U_t|$ \\
      \# enrolled samples & $n$ &  \\
      \# presentation attempts & $c$ &  \\
      \bottomrule
   \end{tabular}
\end{table}

In this section, we formally define the problem and provide a generic formulation of the attack.

\subsection{Speaker Verification Pipeline}
\label{sec:speaker-verification-pipeline}

We consider a standard text-independent speaker verification pipeline where a fixed-length speaker embedding $\mathbf{f} \in \mathbb{R}^{e}$ is extracted from a speech waveform of variable length $\mathbf{w} \in [-1,1]^*$ using a \emph{speaker encoder} $(\mathcal{E})$. Verification for a given user $u$ involves comparison of the speaker embedding $\mathbf{f}$ extracted from a presented test sample with a set of \emph{enrolled embeddings} $F_{u} = \lbrace \mathbf{f}_{\mathtt{u},\mathtt{i}} : \mathtt{u} = u \rbrace$. Without loss of generality, we assume a fixed number of enrolled samples per person ($n$). The number of collected samples and their combination depend on an enrollment and scoring strategy. We discuss each step in detail below. Our notation is shown in Table~\ref{tab:notation}.

\paragraph{Speaker encoder} The speaker encoder $\mathcal{E}$ is typically a DNN trained on an \emph{acoustic representation} $\mathcal{A}(\mathbf{w}) = \mathbf{A} \in \mathbb{R}^{k \times *} $ (e.g., spectrogram or filter banks):
\begin{equation}
   [-1,1]^* \ni \mathbf{w} \xrightarrow{\mathcal{A}} \mathbf{A} \in \mathbb{R}^{k \times *} \xrightarrow{\mathcal{E}} \mathbf{f} \in \mathbb{R}^e
\end{equation}
The model is typically pre-trained for fully supervised closed-world speaker classification on a large corpus with thousands of speakers. Ultimately, the classification head is discarded and the preceding layer is used to extract the \emph{speaker embedding}.

\paragraph{Enrollment and scoring strategy} The system collects multiple speech samples of each user and stores the resulting speaker embeddings, i.e., the database is a collection $F = \lbrace \mathbf{f}_{\mathtt{u},\mathtt{i}} \rbrace$ where $\mathtt{u}$ denotes users and $\mathtt{i}$ indexes their successive samples (for simplicity, assume a constant number of samples $n$ per user). During \emph{verification}, the system returns a binary decision indicating whether a test sample matches a claimed identity $u$. The test waveform $\mathbf{w}$ is encoded analogously to enrollment and processed according to a verification rule:
\begin{equation}
   v_{\rho,\tau}(\mathbf{w}, u | F) := v_{\rho,\tau}(\mathbf{f}, F_u) : \mathbb{R}^e \times \mathbb{R}^{n \times e} \rightarrow \lbrace 0, 1 \rbrace
\end{equation}
which involves the choice of a scoring strategy $\rho$ for combining multiple embeddings/scores and a threshold $\tau$. We consider two popular policies \cite{DBLP:journals/dsp/RajanAHK14}: (1) \emph{any}-$n$ scores similarity with each of the enrolled embeddings and takes the maximum one; (2) \emph{avg}-$n$ scores similarity to the average embedding. Formally:
\begin{align}
  v_{\text{any},\tau} (\mathbf{f}, F_u) & = \text{any} \left( \lbrace \mathbf{f} \circ \mathbf{f}_{u,\mathtt{i}} > \tau : \mathtt{i} = 1, ..., n \rbrace \right) \\
  v_{\text{avg},\tau} (\mathbf{f}, F_u) & = \mathbf{f} \circ \mathbf{f}_u > \tau \\
  & = \mathbf{f} \circ \left( \frac{1}{n} \sum_{\mathtt{i}=1}^{n} \mathbf{f}_{u,\mathtt{i}} \right) > \tau 
\end{align}
where operator $\mathbf{f}_1 \circ \mathbf{f}_2 \rightarrow \mathcal{R}$ denotes a similarity function (e.g., the cosine similarity or the inverse of the Euclidean distance).

\subsection{Dictionary Attack Formulation}

In contrast to classic speaker spoofing, the goal of dictionary attacks is to match a large fraction of an unknown population by pure chance. Formally, the goal of the attacker is to find a master voice sample $\mathbf{w}_*$ that maximizes false matching rate within some user population $U$:
\begin{equation}
   \mathbf{w}_* = \underset{\mathbf{w}}{\operatorname{argmax}} \underset{u \in U}{\operatorname{\mathbb{E}}} \left[ v_{p,\tau}(\mathbf{w}, u) \right]
   \label{eq:master-voice-single-formulation}
\end{equation}

This formulation assumes only a single presentation attempt and is referred to as a \emph{master voice} (MV). However, many verification systems allow for several trials, each possibly using a different utterance. Hence, we distinguish a \emph{maximum coverage master voice} sequence (MCMV) {\color{black} $W_*$}:
\begin{align}
   {\color{black}
   W_* = \underset{(\mathbf{w}_1, \ldots, \mathbf{w}_c )}{\operatorname{argmax}} \underset{u \in U}{\operatorname{\mathbb{E}}} \left[ v_{\rho, \tau}(\mathbf{w}_1, u) \lor \ldots \lor v_{\rho, \tau}(\mathbf{w}_c, u) \right]}
   \label{eq:master-voice-collection-formulation}
\end{align}
This attack is more powerful, since each subsequent attempt can be optimized to target the remaining speaker embedding subspace. See Section~\ref{sec:multi-presentation} for more details. 

\subsection{Attack Implementation}

The proposed attack is untargeted. Its goal is to maximize the \emph{impersonation rate} (IR) in an unseen \emph{test population} $U_t$. {\color{black} The IR is defined as the \emph{expected fraction of user population that can be matched by an attack on an identity verification system}\footnote{{\color{black} Our definition of IR can be seen as a special case of spoof false acceptance rate (SFAR) commonly used in biometric literature and defined as \emph{the number of times an active attack or an impostor is accepted as legitimate divided by the total number of attack or impostor attempts}~\cite{Adler2009}.}}. It can be considered for a single speech sample, or for a sequence of samples crafted for a multi-presentation attack \eqref{eq:master-voice-collection-formulation}. The distinction is clear from context. Formally, given a population $U$ and the corresponding database of enrolled speaker embeddings $F$, the IR of a set of utterances {\color{black}$W$} is estimated as:}
\begin{align}
   {\color{black} 
   \text{IR}(W) = \frac{1}{|U|} \sum_{u \in U} \text{min}\left(1, \sum_{\mathbf w \in W} v_{\rho,\tau}(\mathbf{w}, u | F) \right)}
   \label{eq:ir}
\end{align}

A practical implementation of our attack may use a proxy \emph{optimization population} $\mathcal{U}_o$. Our implementation uses adversarial optimization driven by mean similarity of speaker embeddings in $\mathcal{U}_o$:
\begin{align}
   \mathbf{w}_* & = \underset{\mathbf{w}}{\operatorname{argmax}}~\mathcal{S}(\mathbf{w}, F^o ) \\
   \mathbf{f} & = \mathcal{E}( \mathcal{A} (\mathbf{w})) \\
   ~\mathcal{S}(\mathbf{w}, F^o ) & = ~\frac{1}{n m} \sum_{u \in \mathcal{U}_o} \sum_{i=1}^{n} \mathbf{f} \circ \mathbf{f}_{u,i}
   \label{eq:similarity}
\end{align}
where $m$ is the number of speakers and $F^o$ denotes the gallery of speaker embeddings from the optimization population. As we will demonstrate in the experimental section, the attack transfers between different user populations. Due to observed distinct characteristics of male and female speech (and the resulting remarkable differences in their impersonation susceptibility \cite{DBLP:conf/interspeech/MarrasKMF19}), we focus on attacking a single gender at a time. To avoid unnecessary complexity, we do not reflect this in our notation and simply remark that, unless stated otherwise, we assume and report results for each gender separately.

We show a schematic illustration of the attack in Fig.~\ref{fig:optimization}. The process starts with a \emph{seed} sample ($\mathbf{w}_{0}$) used for initialization of the attack. Then, the speaker encoder $\mathcal{E}$ extracts the embedding $\mathbf{f}_*$  for the optimized sample and its similarity to pre-computed embeddings from the optimization population is calculated as in~\eqref{eq:similarity}. The gradient of the similarity score is then used to iteratively update the chosen representation of the speech sample. Let $\mathtt{t}$ denote the current step and {\color{black} $T$} the total number of steps. The update process has the following general form:
\begin{align}
   \mathbf{v}^{\mathtt{t}+1} & = \mathbf{v}^{\mathtt{t}} + \lambda \nabla_{\mathbf{v}}~\mathcal{S}(\mathcal{G}(\mathbf{w}_{0}, \mathbf{v}^{\mathtt{t}}|\theta), F^o)\\
   \mathbf{w}_* & = \mathcal{G}(\mathbf{w}_{0}, \mathbf{v}^{\color{black}T}|\theta)
\end{align}
where $\mathcal{G}(\mathbf{w}, \mathbf{v}|\theta)$ defines a speech representation/generation function driven by an adversarial \emph{attack vector} $\mathbf{v}$ appropriate for that optimization domain. Our attack is generic and can work with various domains. We experimented with optimization of waveforms, spectrograms, and speaker embeddings (e.g., voice cloning). Each domain has its own peculiarities:

\begin{itemize}
\item \emph{Waveform:} this attack aims to find an adversarial perturbation directly in the waveform domain. 
It starts with a seed sample and has a simple formulation that allows for straightforward inclusion of data augmentation, e.g., via playback simulation (Section~\ref{sec:playback}):
\begin{align}
      \mathcal{G}(\mathbf{w},\mathbf{v}|\theta) & = \mathbf{w} + \mathbf{v}
\end{align}

\item \emph{Spectrogram:} this attack aims to find an adversarial perturbation in the acoustic representation accepted by the speaker encoder as an input. It starts with a seed sample and requires invertibility of the representation to reconstruct the adversarial waveform:

\begin{align}
      \mathcal{G}(\mathbf{w},\mathbf{v}|\theta) & = \mathcal{A}^{-1}(\mathcal{A}(\mathbf{w}) + \mathbf{v})
\end{align}

Practical solutions would short-circuit inversion and feed the distorted acoustic representation to the encoder. Inversion can be performed once at the end. For spectrograms, we used the Griffin-Lim algorithm~\cite{DBLP:conf/icassp/GriffinL83}, but some acoustic representations may not be invertible. 

\item \emph{Speaker embeddings:} this attack finds a voice that maximizes impersonation by optimizing a speaker embedding $\mathbf{f}'$ used for conditioning a speech generation model $\mathcal{C}$: 

\begin{equation}
   \mathcal{C}(\mathbf{f}', \theta) = \mathbb{R}^{e'} \times \mathbb{T}^* \rightarrow \mathbf{w} \in [-1,1]^{*}
\end{equation}

where $\mathcal{C}(\mathbf{f}',\theta)$ denotes a voice cloning system able to generate any utterance (with content specified by a string of tokens $\theta \in \mathbb{T}^*$) spoken by any speaker (specified by a speaker embedding $\mathbf{f}'$ returned by a different encoder $\mathcal{E}'$)~\cite{DBLP:conf/nips/JiaZWWSRCNPLW18,rtvc}. The attack sample generation is:
\begin{equation}
   \mathcal{G}(\mathbf{w},\mathbf{v}|\theta) = \mathcal{C}(\mathbf{v}, \theta)
\end{equation}
In our implementation, we used a seed sample to initialize the embedding, i.e.,  $v^0 = \mathcal{E}'(\mathbf{w}_0)$, but one may skip this step and simply sample a random one instead. 
\end{itemize}

Following the same logic, the attack can be defined in other domains too, e.g., the latent space of a generative adversarial network~\cite{DBLP:conf/icassp/YamamotoSK20}, a (variational) auto-encoder~\cite{DBLP:conf/icassp/WilliamsZCY21}, or the speaker embedding of a voice conversion system~\cite{DBLP:conf/icml/QianZCYH19}.

In practice, we implemented the attack using stochastic gradient descent. Processing the optimization population in batches is both necessary (due to constraints on available GPU memory) and advantageous to the speed of convergence (making more update steps requires fewer epochs). Depending on the size of the speaker encoder, we used batches of 64-256 items. We pre-shuffled the stored embeddings to diversify speaker identities within each batch. Choice of the update step $\lambda$ and the number of epochs depend on the attack configuration. To speed up convergence and parameter choice, we use gradient normalization (with $L_2$ and $L_\infty$ norms). The impact of these parameters is shown in detail in Section~\ref{sec:attack-and-sv-settings}.

\subsection{White-Box vs. Black-Box Attacks}
\label{sec:black-box-intro}

Our attack can be carried out under both the \emph{white-box} and \emph{black-box} threat models. The white-box attack uses gradients computed by automatic differentiation in machine learning frameworks. This allows for fast and accurate optimization but is limited to known models operating in a fully differentiable pipeline. In contrast, the black-box attack uses surrogate gradients estimated by querying any model. This leads to a general attack applicable to all pipelines, but requires larger computational cost. We used the natural evolution strategy (NES)~\cite{DBLP:journals/jmlr/WierstraSGSPS14} to estimate the gradient based on a small number of queries with a Gaussian search distribution with antithetic sampling. {\color{black} Rather than maximizing an objective function directly, NES maximizes the expected value of the objective in the vicinity implicitly defined by a stochastic search distribution. This allows for gradient estimation in fewer queries than typical finite-difference methods.} We leveraged NES as follows:
\begin{align}
   \nabla s(\mathbf{w}) & \approx \nabla \underset{\mathbf{w}'}{\operatorname{\mathbb{E}}}[s(\mathbf{w}')] \approx \frac{1}{ 2 s \sigma} \sum_{i=1}^{s} \delta_i s(\mathbf{w} \pm \sigma \mathbf{\delta}_i) \\
   \mathbf{\delta}_i & \sim \mathcal{N}(0,\mathbf{I}) \\
   s(\mathbf{w}) & = S(\mathbf{w}, F^*) : F^* \sim F^o
\end{align}
where $F^*$ is a batch sampled from the optimization population.

Such an attack requires fine-tuning of two additional hyper-parameters ($s,\sigma$) but can be effective against various models. We discuss this in more detail in Sections~\ref{sec:black-box-evaluation} and \ref{sec:results-cloning}.

\subsection{Playback simulation}
\label{sec:playback}

To assess (and improve) robustness to distortions, we include an optional \emph{playback simulation} step. This step can be included both during evaluation and attack optimization. Let $\circledast$ denote 1D convolution and $(\mathbf{k}_s, \mathbf{k}_r, \mathbf{k}_m)$ denote the impulse responses of the speaker, room, and microphone, respectively. The waveform after playback can be computed as:

\begin{align}
   \mathbf{w}' &= (((\mathbf{w} \circledast \mathbf{k}_s) + \mathbf{n}) \circledast \mathbf{k}_r) \circledast \mathbf{k}_m \\
   \mathbf{n} & \sim \mathcal{N}(0, \alpha \mathbf{I})
\end{align}

To increase augmentation diversity, we randomize the simulation by sampling the AWGN strength $\sqrt{\alpha} \sim \mathcal{N}(0, 0.025)$ and choosing random kernel combinations from a small database with 4 speakers, 9 rooms and 7 microphones.

\subsection{Multiple Presentation and Coverage Optimization}
\label{sec:multi-presentation}

Due to their imprecise nature, biometric systems often allow for a few authentication attempts before falling back to a PIN or passphrase. This behavior can be exploited and the attacker can craft a sequence of diverse speech samples that maximize the overall success rate. Let $c$ denote the number of allowed attempts. The attacker can simply generate a set of master voices $\lbrace \mathbf{w}_\mathtt{c} : \mathtt{c} = 1, \ldots, c' \rbrace$, optimized independently as in Eq. \eqref{eq:master-voice-single-formulation} based on $c' \gg c$ randomly chosen seed samples. Finally, the best $c$ samples ($\mathbf{w}_{\mathtt{c}_1}, \ldots, \mathbf{w}_{\mathtt{c}_c}$) are chosen for the attack. 

In our experiments, we simply reuse the \emph{optimization} population $U_o$ to assess viability of candidate samples. The attacker computes $\mathbf{B} = [b_{\mathtt{c},\mathtt{u}}]$, a binary \emph{impersonation matrix} indicating matching success for the $\mathtt{c}$-th sample against user $u \in U_o$:
\begin{equation*}
   \mathbf{B}_{c' \times m} = 
   \begin{bmatrix}
      v_{\rho, \tau}(\mathbf{w}_1, u_1) & \ldots & v_{\rho, \tau}(\mathbf{w}_1, u_m) \\
      \ldots & \ldots & \ldots \\
      v_{\rho, \tau}(\mathbf{w}_{c'}, u_1) & \ldots & v_{\rho, \tau}(\mathbf{w}_{c'}, u_m) \\
      \end{bmatrix}
\end{equation*}

Aggregation along the user dimension yields expected IRs. We hence test two simple strategies:

\begin{itemize}
   \item naive \emph{independent selection} takes the top-$c$ speech samples based on their IR on the entire optimization population, i.e., the $i$-th sample is simply:
   \begin{equation}
      \mathbf{w}_{\mathtt{c}_i} : \mathtt{c}_i = \underset{\mathtt{c} \notin \lbrace \mathtt{c}_1, ..., \mathtt{c}_{i-1} \rbrace }{\operatorname{\text{argmax}}} \frac{1}{m} \sum_{u \in U_o} b_{\mathtt{c}, u}
   \end{equation}   
   
   \item \emph{complementary selection} takes a single best sample step-by-step, each time maximizing the IR on the still \emph{uncovered subset} of the optimization population, i.e., the $i$-th sample is chosen as:
   \begin{equation}
      \mathbf{w}_{\mathtt{c}_i} : \mathtt{c}_i = \underset{\mathtt{c} \notin \lbrace \mathtt{c}_0, ..., \mathtt{c}_{i-1} \rbrace }{\operatorname{\text{argmax}}} \frac{1}{|U_i|} \sum_{u \in U_i} b_{\mathtt{c},u}
   \end{equation}
   \begin{equation}      
      U_i = 
      \begin{cases}
         U_o & \text{for } i = 1 \\
         U_{i-1} \setminus \lbrace \mathtt{u} : v_{\rho,\tau}(\mathbf{w}_{\mathtt{c}_{i-1}}, \mathtt{u}) \rbrace & \text{for } i > 1
      \end{cases}
   \end{equation}
\end{itemize}

An interesting extension of this problem would be to jointly optimize all $C$ samples. We leave this aspect for future work.

\section{Experimental Setup}

In this section, we show our experimental setup and the details of the used datasets, speaker encoders etc. We explain our model pre-training and calibration, and provide an exhaustive benchmark evaluation of the resulting speaker verification systems, including key aspects of their menagerie analysis.

\subsection{Datasets}

We used two public datasets in our work: VoxCeleb~\cite{DBLP:journals/csl/NagraniCXZ20} and LibriSpeech~\cite{DBLP:conf/icassp/PanayotovCPK15}. VoxCeleb~\cite{DBLP:journals/csl/NagraniCXZ20} is a large, state-of-the-art dataset of human speech composed of two parts:  VoxCeleb1 (dev set: 1,221 speakers and 148,642 utterances; test set: 40 speakers and 4,874 utterances) and VoxCeleb2 (dev set: 5,994 speakers and 1,092,009 utterances; test set: 119 speakers and 36,237 utterances). Both parts are fairly gender-balanced (55\% and 61\% of male speakers, respectively) and feature speakers from various ethnicities, accents, and age groups. 
Original videos used for speech extraction were shot in a wide range of challenging environments, including red carpet interviews, outdoor stadiums, indoor studios, speeches given to large audiences, excerpts from professionally shot multimedia, and amateur footage shot on hand-held devices.
Crucially, they represent challenging real-world conditions which vary in background chatter, room acoustics, overlapping speech, recording equipment quality, and surrounding noise.

\begin{table}[!b]
\caption{Dataset partition.}
\label{tab:data}
\resizebox{\columnwidth}{!}{
\begin{tabular}{lllrrr}
\toprule
\textbf{Dataset}               & \textbf{Partition} & \textbf{Scope} & \multicolumn{1}{r}{\textbf{$|U|^1$}} & \multicolumn{1}{r}{\textbf{$|A|^2$}} & \multicolumn{1}{r}{\textbf{$|A| / |U|$}} \\
\midrule
VoxCeleb1-Dev & P1      & SV Train             & 1,211                            & 148,642                          & 122                                    \\
VoxCeleb1-Test & P2   & SV Eval.             & 40                               & 4,874                            & 122                                    \\
VoxCeleb2-Dev & P3a    & MV Opt.       & 1,000                            & 50,000                           & 50                                     \\
                               & P3b  & MV Eval.           & 1,000                            & 100,000                          & 100                                    \\
                               & P3c  & SV Train            & 3,994                            & 895,664                          & 224   \\
\bottomrule

\multicolumn{5}{l}{\color{black} $^1$ $|U|$ refers to the number of included users in total} \tabularnewline
\multicolumn{5}{l}{\color{black} $^2$ $|A|$ refers to the number of included utterances in total}
\end{tabular}
}
\end{table}

\begin{table}[!t]
\caption{Speaker encoder models and their benchmark performance}
\label{tab:encoders-compact}
\resizebox{\columnwidth}{!}{\input{tables/encoders_compact.tex}}
\end{table}

We divided the VoxCeleb speakers into disjoint partitions (Table~\ref{tab:data}). First, we sampled two gender-balanced subsets of 1,000 people for master voice optimization and testing (partitions P3a and P3b, respectively). The remaining 5,205 speakers (P1 and P3c) were used for speaker encoder training. For consistency with standard evaluation methodology, we used the VoxCeleb1 test partition (P2) for speaker encoder benchmarking and calibration. In contrast, LibriSpeech is a clean and transcribed dataset with high-quality recordings of English speakers reading excerpts from audio books in studio conditions, used only in our final voice cloning experiments.

In all experiments, we use single-channel 16-bit audio with 16~kHz sampling rate. The waveforms are normalized to [0,1] and standardized to be 2.58~second long, which is achieved by random cropping (mostly) or zero-padding (occasionally). 

\subsection{Speaker Encoders}

We used speaker encoders based on various CNN backbones and acoustic representations, including adapted VGG (VGGVox~\cite{DBLP:journals/csl/NagraniCXZ20}) and ResNet models (ResNet 50~\cite{DBLP:journals/csl/NagraniCXZ20} and Thin ResNet~\cite{DBLP:conf/icassp/XieNCZ19}) trained on spectrograms, and x-vector based on filter banks~\cite{DBLP:conf/icassp/SnyderGSPK18}. 
VGG and ResNet models were adapted from computer vision to spectrogram inputs by replacing the last fully-connected (FC) layer with two layers: a FC one with support in the frequency domain and average pooling with support on the time domain.
X-vector~\cite{DBLP:conf/interspeech/SnyderGPK17} is a TDNN, which allows neurons to receive signals spanning multiple frames. Given a filter bank, the first five layers operate on speech frames, with a time context centered at the current frame. 
A pooling layer aggregates frame-level outputs and computes mean and standard deviation. 
Two FC layers aggregate statistics across the time dimension. We used a GhostVLAD pooling layer~\cite{DBLP:conf/icassp/XieNCZ19}, with 10 clusters plus 2 ghost clusters, on all models. 

\subsection{Pre-processing and Training}

We trained our speaker encoders from scratch using samples from 5,205 speakers (partitions P1 and P3c). 
We randomly sampled segments from each of their utterances and standardized the inputs to 2-second clips (by cropping or padding, respectively). No voice activity detection or silence removal was applied. 
Spectrograms (filter banks) were generated in a sliding window fashion using a Hamming window of width 25ms and step 10ms. 
We used 512-point (Fast Fourier Transforms) FFTs yielding spectograms of size 257$\times$200 and filter banks of size 24$\times$300 (frequency$\times$temporal). Each acoustic representation was normalized by subtracting the mean and dividing by the standard deviation of all frequency components in a single time step. The model was trained for classification using Softmax and the Adam optimizer, with an initial learning rate of 0.001, decreased by a factor of 10 after every 10 epochs. 

\subsection{Speaker Verification Performance}

\begin{figure*}[!t]
\centering
\subfloat[VGGVox]{\includegraphics[width=0.24\textwidth]{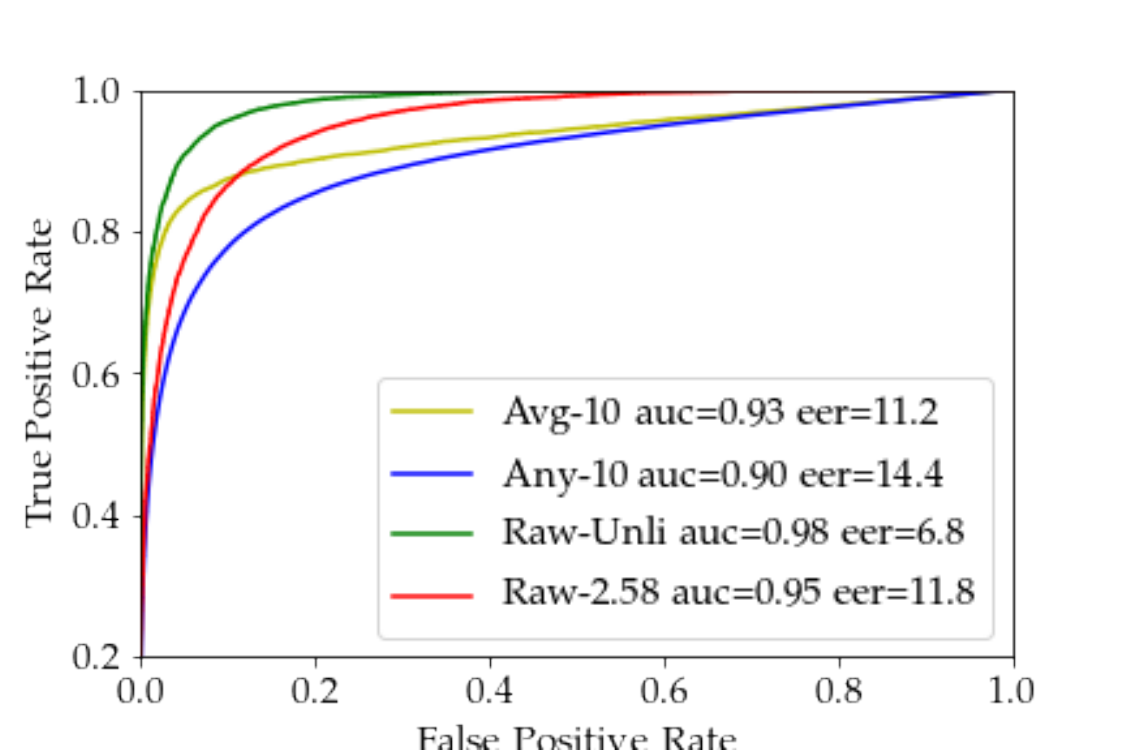}}
~
\subfloat[ResNet 50]{\includegraphics[width=0.24\textwidth]{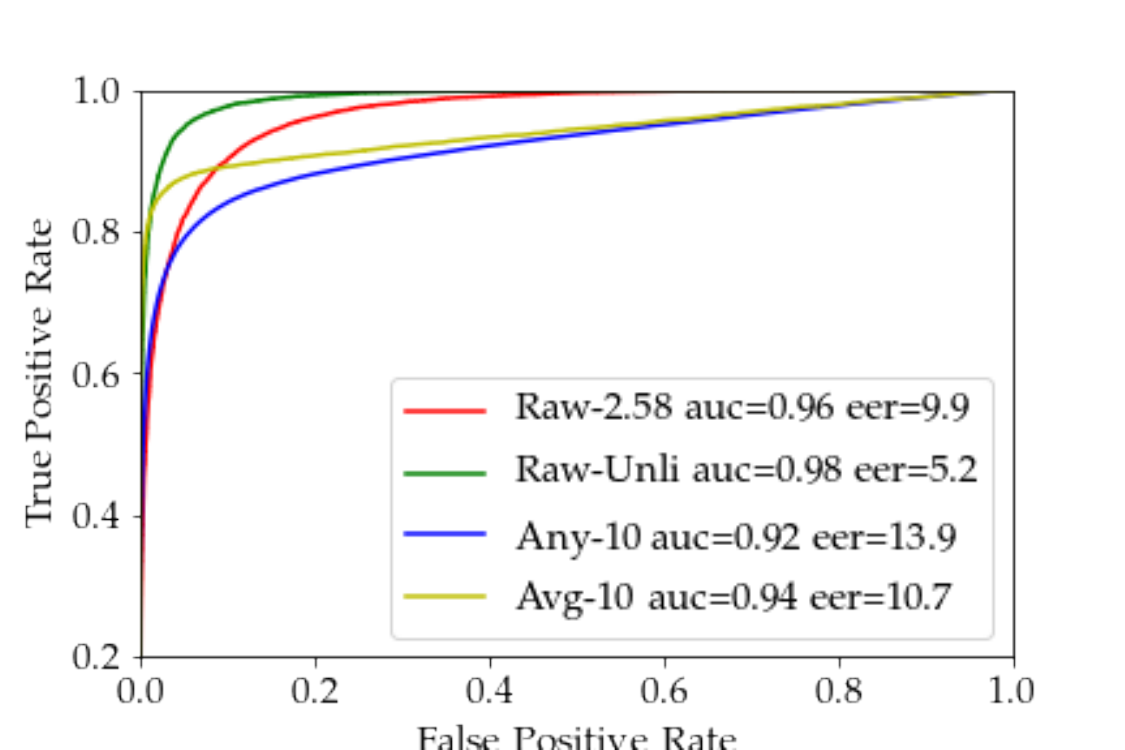}}
~
\subfloat[Thin ResNet]{\includegraphics[width=0.24\textwidth]{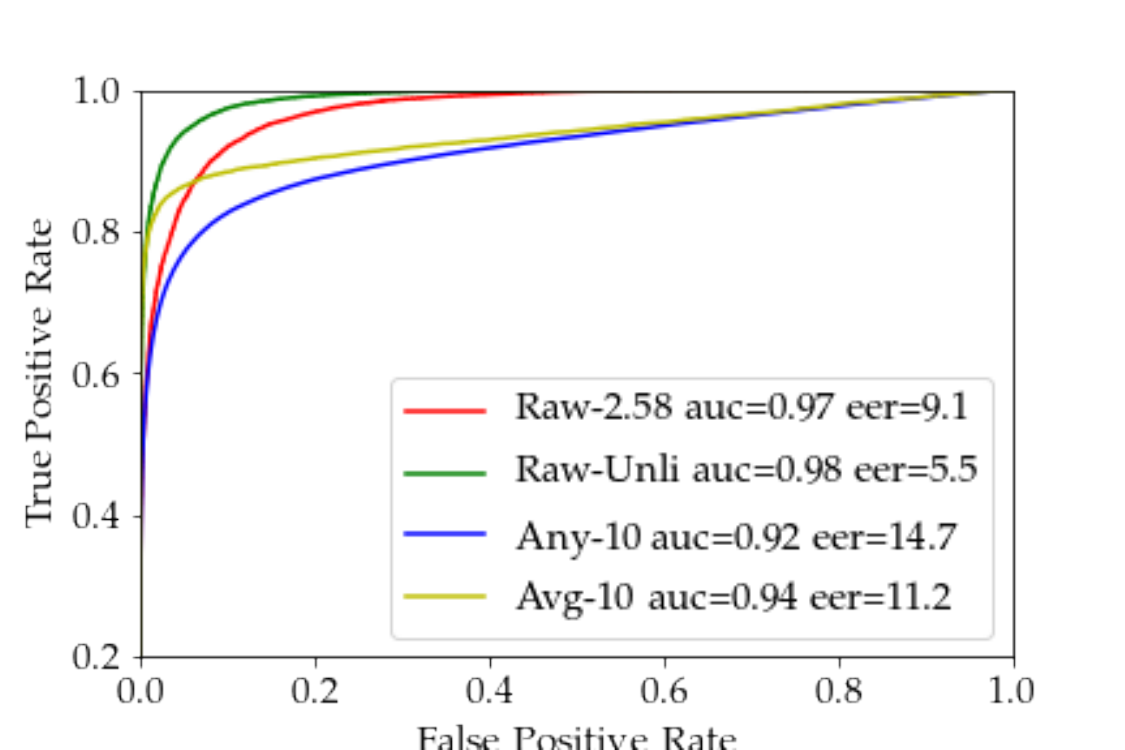}}
~
\subfloat[X-vector]{\includegraphics[width=0.24\textwidth]{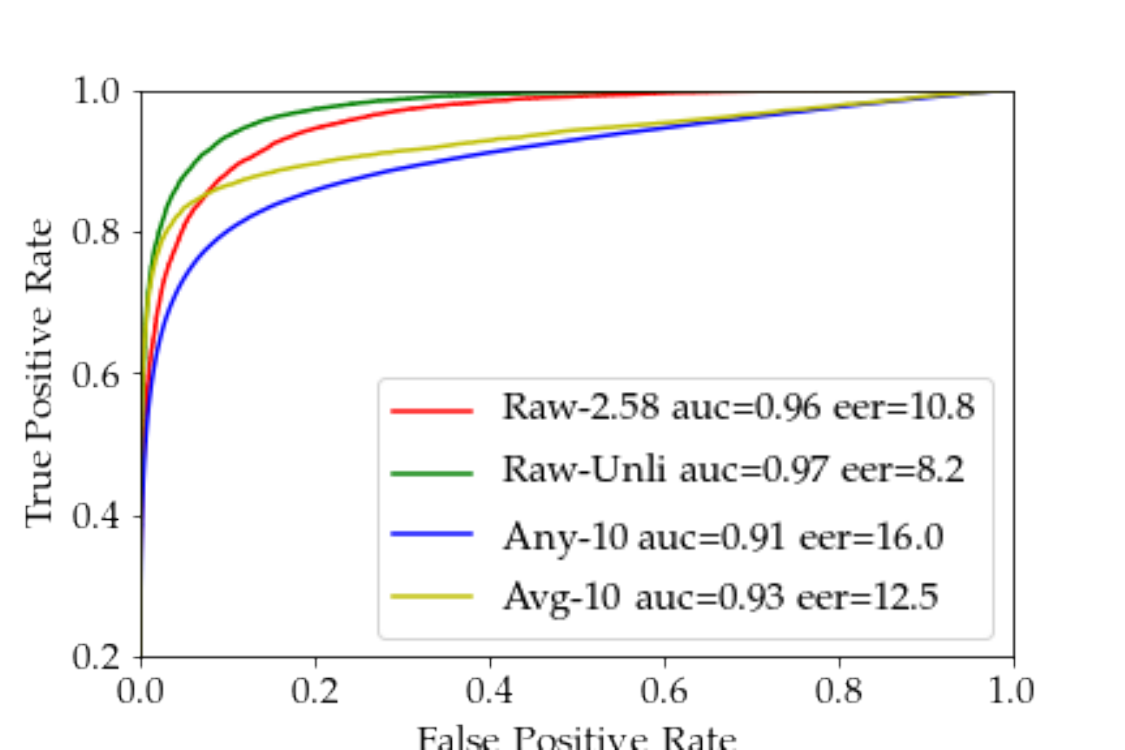}}
\caption{Receiver operation characteristics of the considered speaker encoders. We include both raw discriminability of the embeddings and final performance accounting for the enrollment and verification policy. It can be observed that the \emph{avg} policy leads to a more secure system in the low FAR regime.}
\label{fig:ffdb}
\end{figure*}

The pre-trained speaker encoders are deployed in a speaker verification system (Section~\ref{sec:speaker-verification-pipeline}) by stripping their classification heads. We performed a detailed assessment of open-set speaker verification performance that includes two key aspects: 

\begin{itemize}
   \item \emph{Raw performance:} we test discriminability of speaker embeddings on the standard VoxCeleb1 test pairs (37,720 pairs; partition P2). Based on the collected cosine similarities, we find the ROC and derive common metrics, e.g., area under the curve (AUC) and equal error rate (EER). 
   \item \emph{Deployment performance:} we test performance that accounts for the enrollment and scoring strategy. We used a larger population of 1,000 people (partition P3a). 
\end{itemize}
The resulting evaluation will be used for threshold calibration - to summarize the behavior we focus on thresholds corresponding to the EER and a 1\% false acceptance rate (FAR-1)\footnote{
\color{black} False Rejection Rate (FRR) and False Acceptance Rate (FAR) are often referred to as “miss” and “false alarm rates” in speaker verification literature.
}. 
The obtained results are summarized in Table~\ref{tab:encoders-compact} (extended version is reported in Table~\ref{tab:encoders} and the corresponding ROC curves are shown in Fig.~\ref{fig:ffdb}). To enable comparison with related work and investigate the potential gap in typical deployments, we distinguish between evaluation on full audio clips (as performed in the literature) and short utterances only (as used in real deployments). For the latter, we randomly crop a 2.58~second segment. On full-length clips from the standard test set, our models reach EER of $\approx$5-8\% - higher than the best reported results ($\approx$2.5-5\%), but reasonable given our substantially smaller training population\footnote{\color{black}
Note that in contrast to the standard practice in speaker verification literature, we excluded 2,000 people from the training set ($\approx$30\% of the training population) for our master voice analysis. In our preliminary experiments we performed sanity checks on the entire population ($\approx7,200$ people) and were able to obtain EERs within 0.8\% (percentage points) of the results in \cite{DBLP:journals/csl/NagraniCXZ20}.
}. On shorter clips, this deteriorates across all models down to $\approx$9-11\%.

Enrollment of multiple samples and using a scoring strategy (tested on short clips only) can substantially improve performance, especially in the low FAR regime. We observed the best results with the \emph{avg}-$10$ policy, which is consistent with earlier findings~\cite{DBLP:journals/dsp/RajanAHK14}. The \emph{any}-$10$ policy was consistently inferior to the use of even a single speaker embedding.

\subsection{Dictionary Attack Implementation Details}
\begin{figure*}[!t]
\centering
\subfloat[VGGVox\label{a_vggvox}]{
   \includegraphics[width=0.24\linewidth]{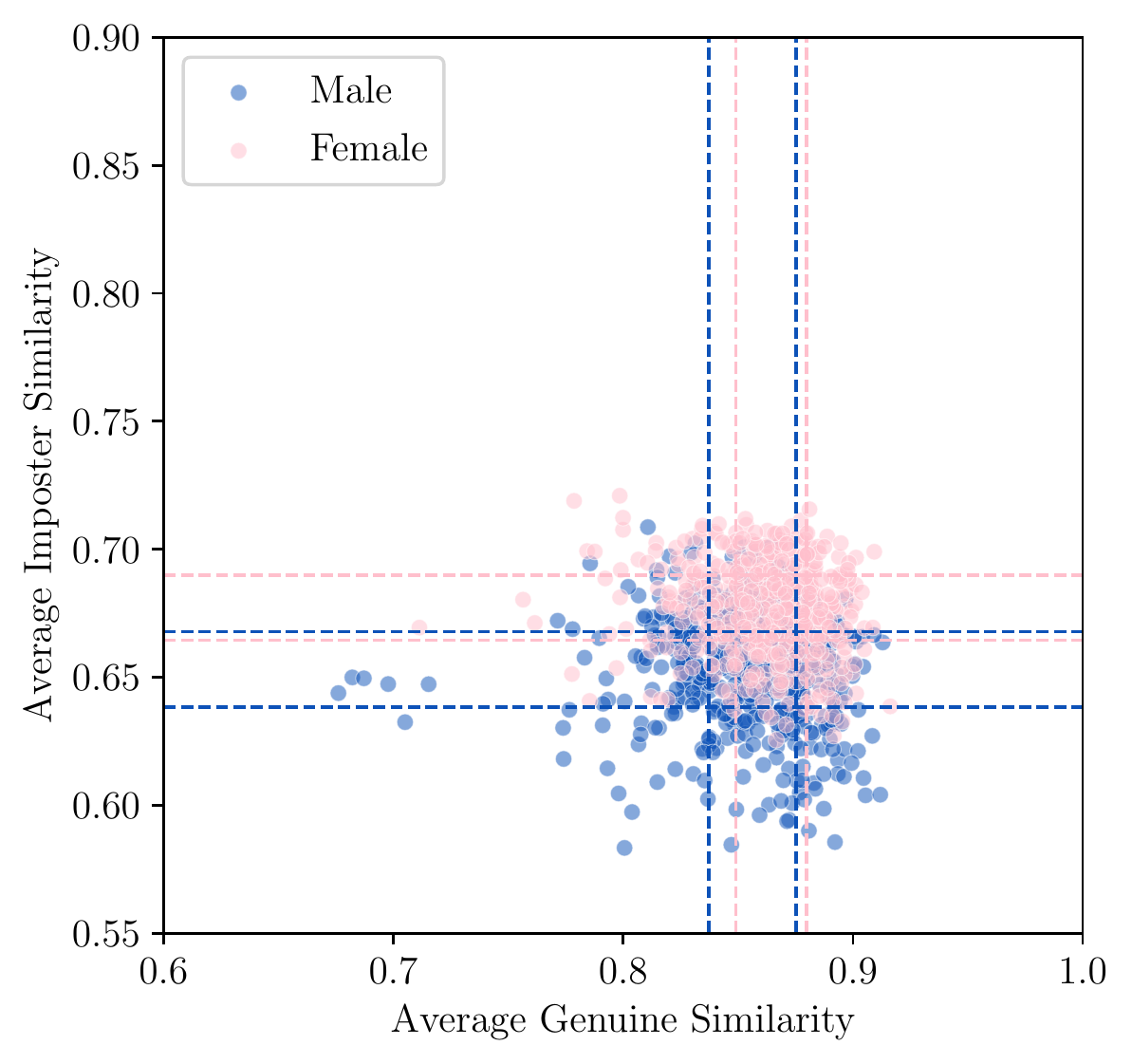}}
\subfloat[ResNet 50\label{a_resnet50}]{
   \includegraphics[width=0.24\linewidth]{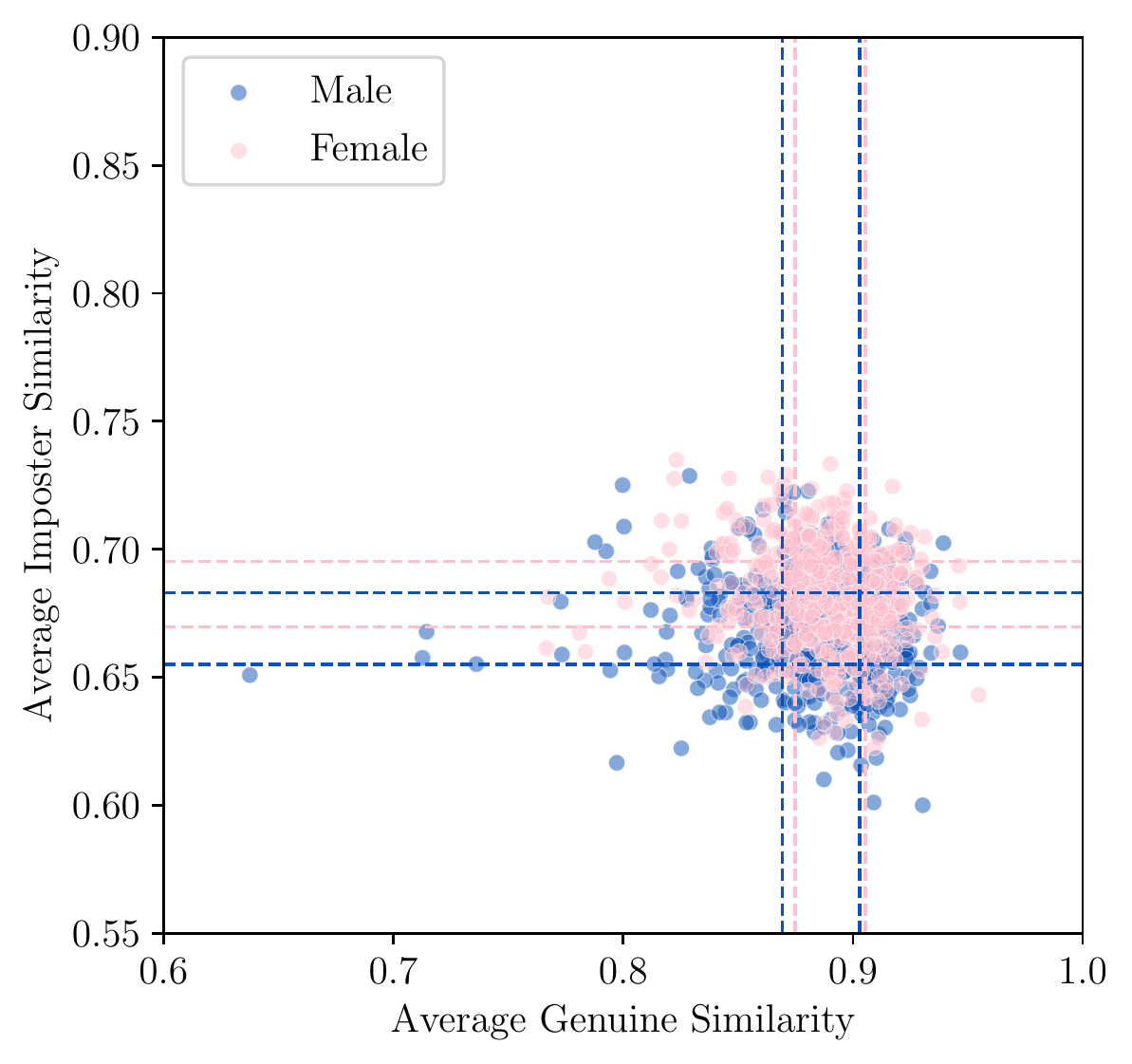}}
\subfloat[Thin ResNet\label{a_thinresnet}]{
   \includegraphics[width=0.24\linewidth]{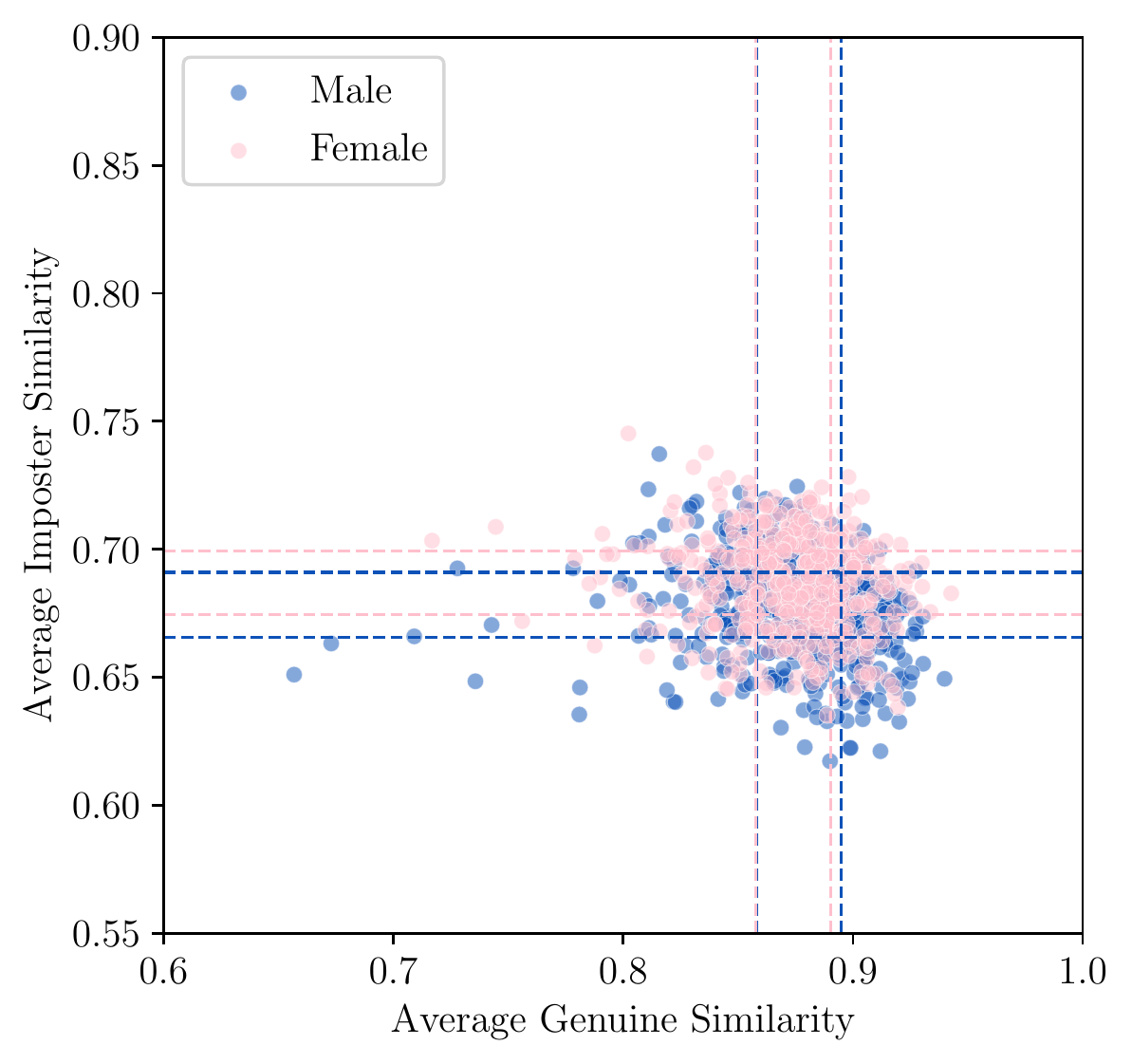}}
\subfloat[X-vector\label{a_xvector}]{
   \includegraphics[width=0.24\linewidth]{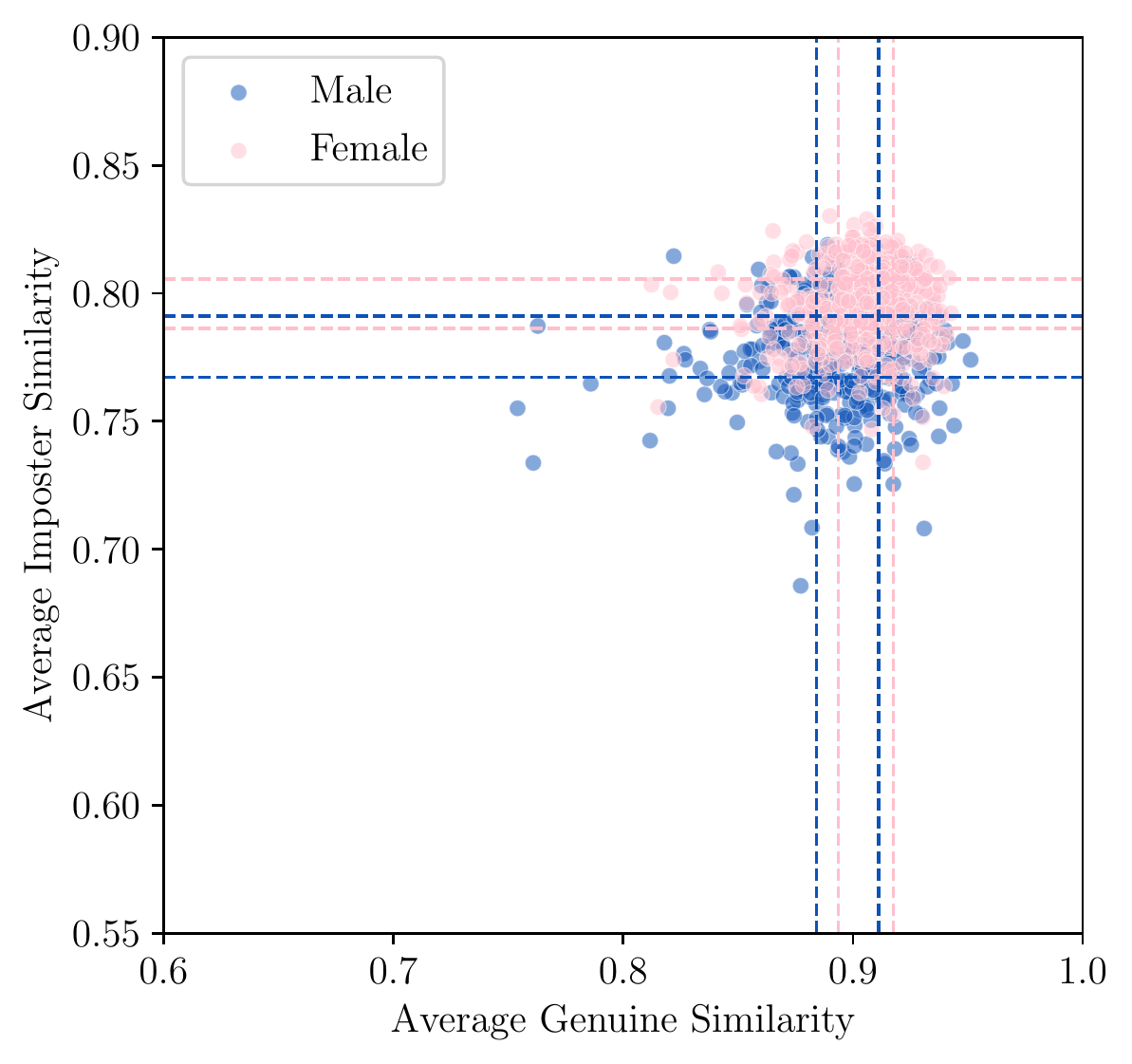}}
\caption{For each speaker encoder, a menagerie analysis plot. Each point in a plot represents a user, defined by their own average imposter score (when matched with other people) and the average genuine score (when matched with others of their own examples). For both the x- and y-axis, the two dashed lines indicate the 25\% and 75\% percentile. It should be noted that female users tend to have a higher average imposter score. The range of similarity scores is small, therefore even small differences in these average scores can determine highly important differences in impersonation rates between the genders. 
{\color{black} An ideal speaker recognition system should locate users at the lower right corner. Ideal impersonators would be located in the top part of the plot.} }
\label{fig:exp-menagerie}
\end{figure*}

\begin{figure*}[!t]
\centering
\subfloat[VGGVox\label{b_vggvox}]{
   \includegraphics[width=0.24\linewidth]{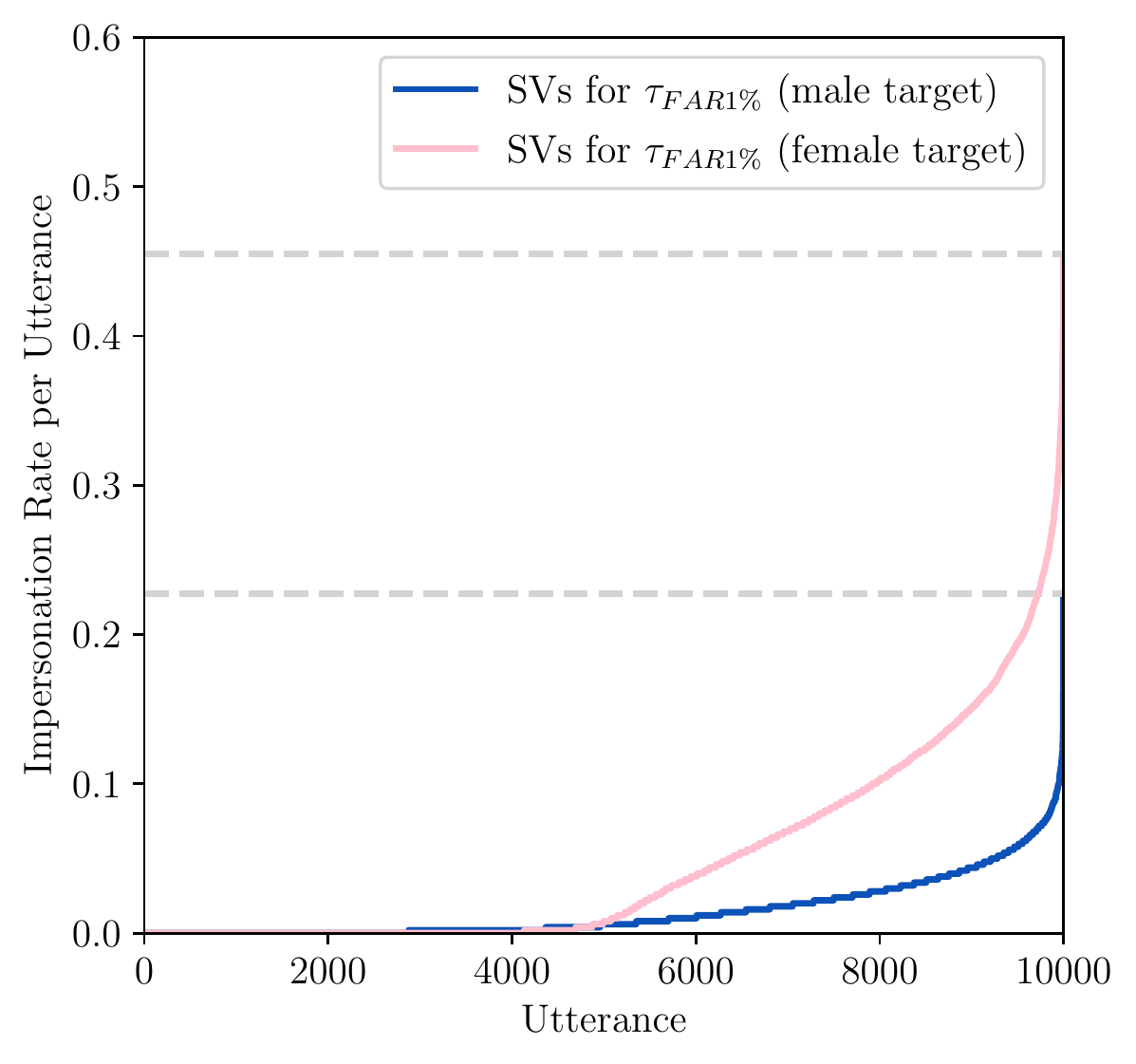}}
\subfloat[ResNet 50\label{b_resnet50}]{
   \includegraphics[width=0.24\linewidth]{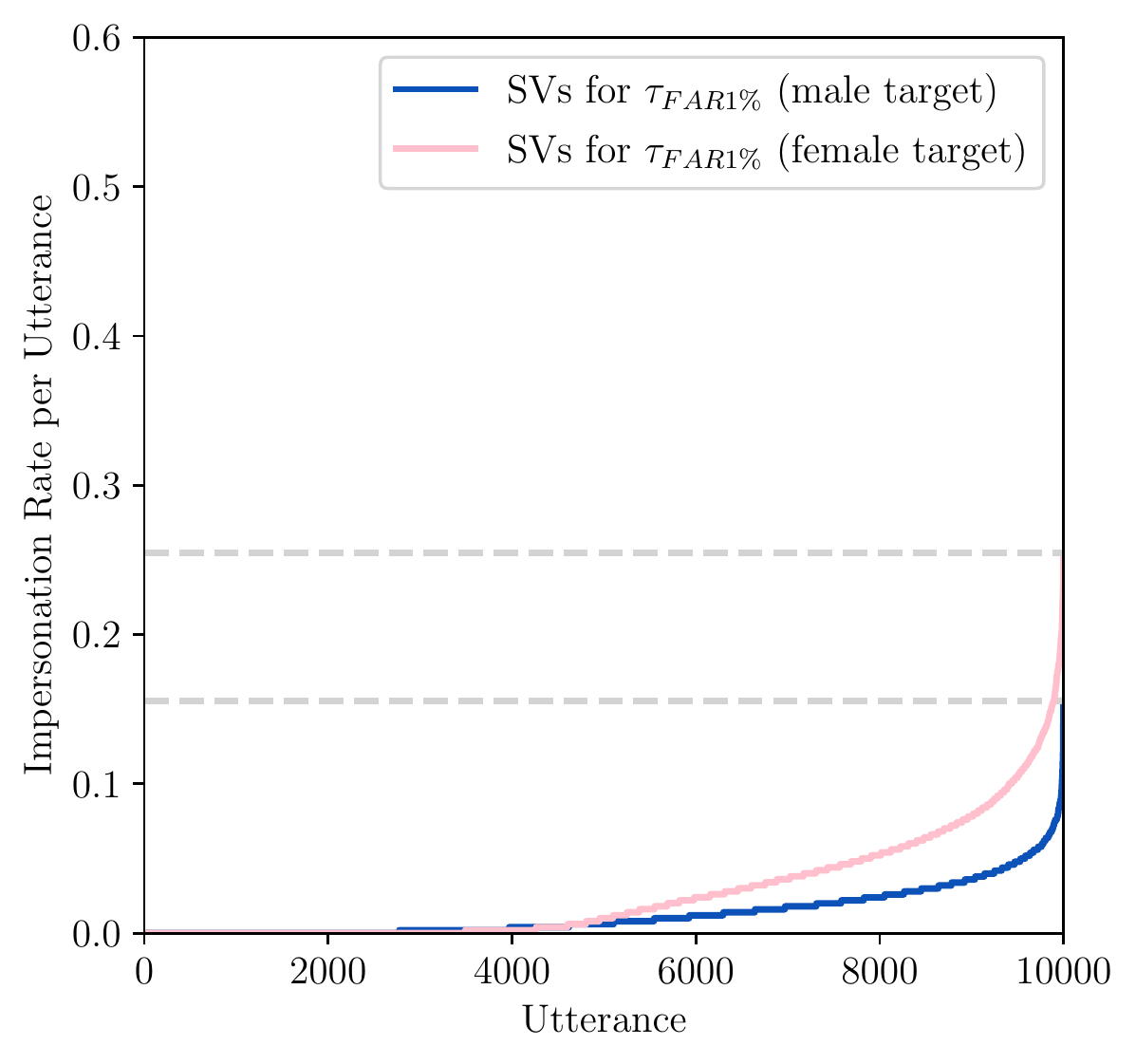}}
\subfloat[Thin ResNet\label{b_thinresnet}]{
   \includegraphics[width=0.24\linewidth]{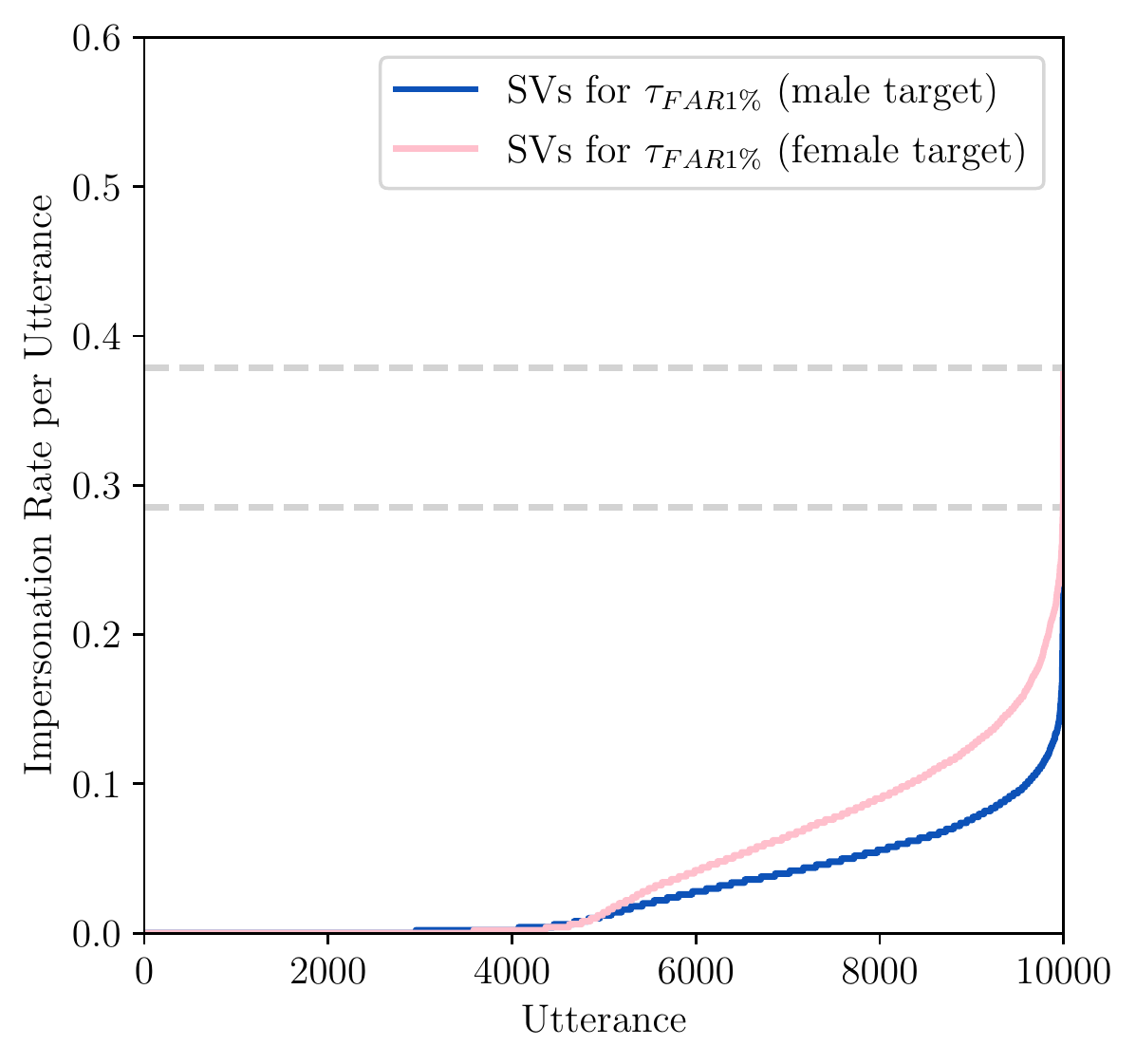}}
\subfloat[X-vector\label{b_xvector}]{
   \includegraphics[width=0.24\linewidth]{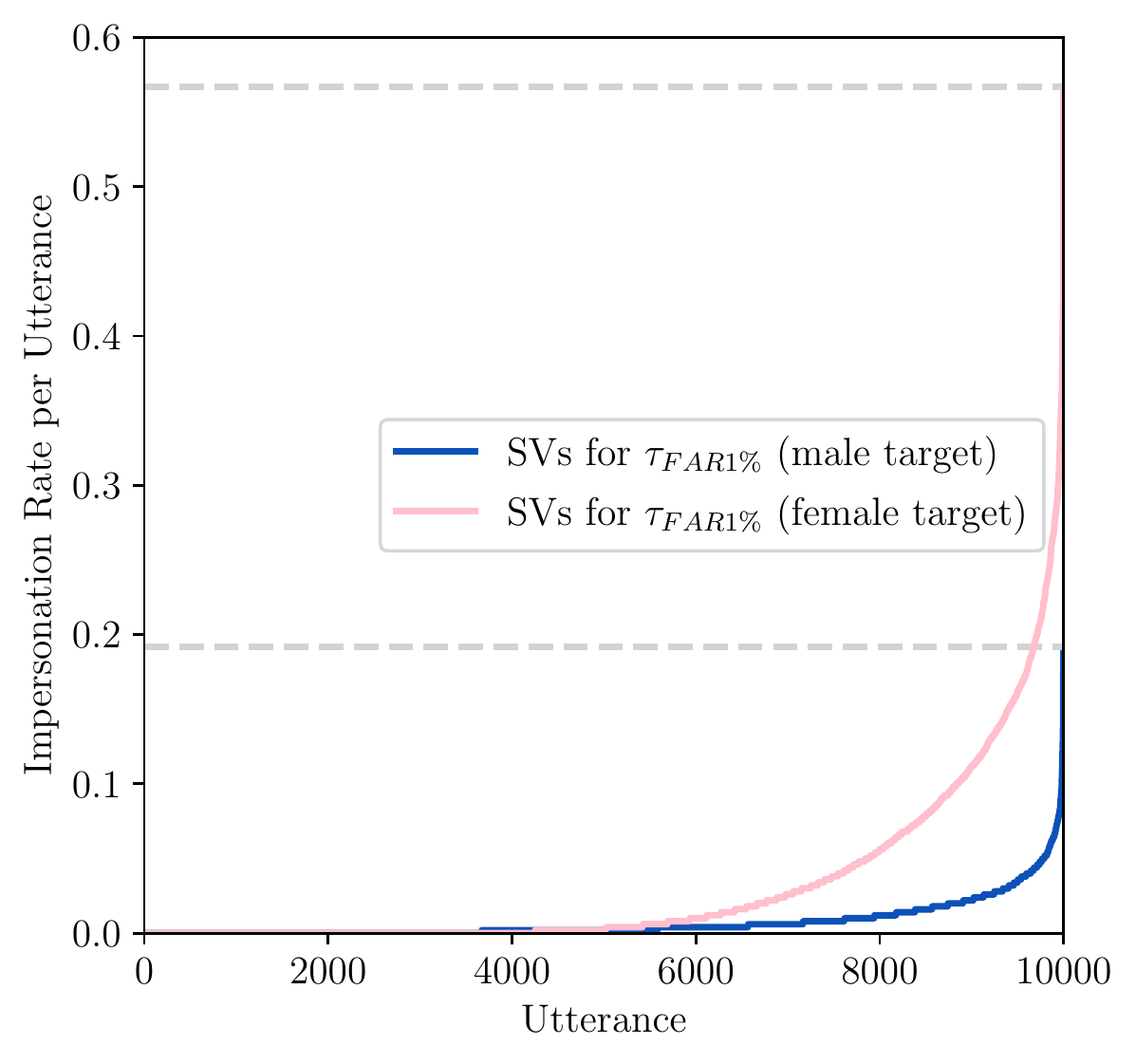}}
\caption{For each speaker encoder, a ranking of all seed utterances according to their Impersonation Rate (IRs) against male (blue) and female (pink) users, under \emph{avg}-10 enrolment/verification policy with a raw \emph{far}-1 threshold. Utterances are sorted by left to right based on an increasing IR. The higher the IRs, the more the utterance tends to match users. It should be noted that female users tend to be impersonated more. VGGVox and x-vector exhibit the larger difference in the maximum IR between genders and represent the least secure system.}
\label{fig:exp-rankings}
\vspace{-4mm}
\end{figure*}

We rely on two disjoint populations for master voice optimization (partition P3a) and testing (P3b). Each population contains 1,000 speakers. We treat male and female speakers separately since their speech exhibits distinct properties and leads to differences in verification performance and vulnerability to impersonation attacks. Specifically, a menagerie analysis on the seed utterances included in partition P3a, whose details are reported in Fig. \ref{fig:exp-menagerie}, showed that women often have a higher average imposter score. This gender-wise difference is emphasized when we consider the impersonation rates achieved by the same seed voices against users from the two genders. Fig. \ref{fig:exp-rankings} shows that women tend to be impersonated more, even under the most secure setting (\emph{avg}-10, raw \emph{far}-1 threshold). VGGVox and x-vector are the least secure systems and exhibit the largest difference in the maximum IR between genders.

The optimization process starts with a seed sample. We randomly sampled 100 seeds for both male and female speakers from P3a. While this step could also possibly be exploited to further improve IRs, we opted for fully random selection to simplify the experiments and rely on a single set of seed voices regardless of the target model. The way seed voices are used differs among the attacks. When optimizing waveforms or acoustic representations, we start with the full content of a seed sample for the target gender. For other attacks, e.g., based on voice cloning, we used seed samples to initialize speaker embeddings that condition the generator. Based on the adopted representation and capabilities of the synthesis model, seed samples may not be needed.

During the attack, the speaker encoder operates in a configuration which compares the current attack sample with a batch of samples from the optimization population. We shuffle samples from various users to promote speaker diversity within each batch. After each batch, we normalize the gradients\footnote{\color{black}
To control the change in magnitude for the gradients and facilitate the selection of the learning rate, 
we divided the gradients by their $L_2$ norm. 
}. and apply the update. Using stochastic gradient descent is beneficial and leads to remarkably reduce optimization time (fewer passes over the entire population, compared with gradient accumulation). We used batches of 64-256 samples, based on the model size and GPU memory (single RTX 8000 GPU). 

To monitor, we track IRs for a single scoring strategy after each epoch (\emph{any}-$10$, raw \emph{far}-1 threshold). At this stage, we stay with the adopted representation. If applicable, we return to the waveform domain (e.g., Griffin-Lim inversion~\cite{DBLP:conf/icassp/GriffinL83}) after optimization ends. We then test speaker verification performance and compare IRs for seed-master voice samples. We initially compare various enrollment policies and decision thresholds, but then focus on one representative configuration.

\input{section_results.tex}

\section{Conclusions and Future Work}

In this paper, we proposed dictionary attacks against speaker verification, a novel attack vector which aims to match a large fraction of user population by pure chance. In contrast to well known spoofing attacks that target one specific individual, our attack aims to exploit the biometric menagerie property - an inherent diversity in matching propensity/susceptibility across different people. Our approach is general and can be applied in various domains, including waveforms, acoustic representations, or even speaker embeddings in voice synthesis systems. We tested several different speaker encoder architectures and considered both white-box and black-box threat models.

We performed the first comprehensive evaluation of dictionary attacks against deep learning based speaker verification systems. The key conclusions from our work are as follows:
\begin{enumerate}
   \item Speech appears to be susceptible to dictionary attacks. We were able to consistently and substantially increase IRs for all considered speaker encoders. Even for the most restrictive threshold (\emph{far}-1 calibrated for the \emph{avg}-$10$ scoring strategy), we were able to craft adversarial waveforms matching 69\% of females and 38\% of males in a population of 1,000 people (Table~\ref{tab:enrollment-policy-breakdown}).
   \item Susceptibility to the attack can vary remarkably across genders. We consistently observed much larger IRs for female speakers. The cause of this discrepancy is not clear. Our training set was only slightly imbalanced (64\% of male speakers) and recent studies found only weak impact of gender balance on the overall error rates even for more unbalanced settings~\cite{fenu2021fair}. Further investigation of this aspect is needed. 
   \item Adversarial optimization driven by raw embedding similarity on a proxy population is a simple and effective attack strategy (e.g., it does not depend on configuration details, such as enrollment policy or decision threshold). The attack works well across speech representations (waveform, spectrogram, speaker embedding) in white- and black-box threat models. In a challenging scenario, our black box attack based on NES was highly effective even when targeting a complex black-box voice cloning system with highly variable output (Section~\ref{sec:results-cloning}).
   \item Our attack transfers across populations but not necessarily across genders. No notable differences in IRs between test and optimization populations were observed. Male and female speech tend to have different characteristics, and targeting both appears to be ineffective. We got best results when seed and target genders match.
   \item Choice of speech representation has crucial impact on the attack. Optimization in the waveform and spectrogram domains leads to adversarial samples with crafted noise. Despite being very effective against the targeted model, it does not transfer between encoder architectures. Optimization of the speaker embedding in voice cloning led to a less effective but transferable attack.
\end{enumerate}

Our results show that dictionary attacks could be a serious threat to speaker verification. {\color{black}We suspect there are two main factors at play. First, the speaker embedding space is likely not distributed uniformly. Regions of high and low density manifest themselves as differences in matching propensity/susceptibility which are characterized via the biometric menagerie. Our attack can take advantage of modern optimization methods and generative models to find speech properties that exploit this property. The higher transferability obtained through voice cloning suggests the existence of high-level master voice characteristics. Secondly, it appears that speaker encoders lack adversarial robustness and allow for finding noise-like perturbations that can maximize similarity even further.}

That being said, our work has several limitations and should be seen as the first step in this direction. {\color{black}We designed our study to include both various speaker encoders and acoustic representations and to evaluate them fairly under the same conditions. Nevertheless, the state-of-the-art in both speaker verification and speech synthesis is moving quickly and more work will be needed to consider both classic approaches (e.g., GMM-UBM~\cite{DBLP:journals/dsp/ReynoldsQD00} or i-vector~\cite{DBLP:journals/taslp/DehakKDDO11}) and emerging architectures (e.g., TDNN \cite{desplanques2020ecapa} or s-vectors based on transformers~\cite{DBLP:journals/corr/abs-2008-04659}).}

The second main limitation of our current attack is the need for validation in a real over-the-air setting. So far, we relied on playback simulation which reveals that the signal distortion introduced by the channel can most likely be dealt with by means of augmented training. However, attacking real deployments will also need to address other factors, e.g., temporal shifts stemming from unknown start and duration of the sample. Another limitation is that both the feasibility of the attack and applicability of various potential countermeasures will come down to the root cause of the vulnerability. While \emph{noise-based adversarial samples} will be difficult to use in practice (e.g., due to unknown model architecture or the need to perform real-time adversarial optimization in a challenge-response regime), \emph{identity-based samples} could potentially scale quite easily (e.g., with pre-computed master embeddings used in a real time voice conversion system). Given the observed transferability of master voice samples obtained with voice cloning, it will be exciting to explore other speech representations and generators, such as disentangled representations~\cite{DBLP:conf/icassp/WilliamsZCY21} and voice conversion systems~\cite{DBLP:conf/icml/QianZCYH19}. 

If our attack will become a viable threat vector, further work will be needed to devise countermeasures (e.g., by deploying an additional background population that reveals mass impersonation capabilities of the given sample). Overall, we believe our work will ultimately lead to a better understanding of the speech modality and more secure human-computer interaction.

\printbibliography
\input{bios/marras_bio.tex}

\vskip -0.5\baselineskip plus -1fil 

\input{bios/korus_bio.tex}

\vskip -0.5\baselineskip plus -1fil 

\input{bios/jain_bio.tex}

\vskip -0.5\baselineskip plus -1fil 

\input{bios/memon_bio.tex}

\vskip -0.5\baselineskip plus -1fil 

\appendices
\counterwithin{figure}{section}
\counterwithin{table}{section}

\section{}
To better understand the context of our study, this appendix collects a range of supplementary results and material:
\begin{itemize}
\item Detailed benchmark performance of the considered speaker encoders at different security thresholds.  
\item Receiver operation characteristics (ROC) curve for the speaker encoders considered in our study. 
\item Menagerie analysis conducted under each of the considered speaker encoders. 
\item Seed utterances ranking based on their impersonation rate against male and female users.
\item The susceptibility of a utterance to achieve high impersonation rates between two speaker encoders.
\item Impersonation rates of seed and master voices in case of multiple presentation attempts for NES-based attacks.
\item Source code and data accompanying this paper available at \url{https://github.com/mirkomarras/dl-master-voices}.
\end{itemize}

\begin{table*}[!t]
\caption{Extended benchmark performance for speaker encoders}
\label{tab:encoders}
\resizebox{\textwidth}{!}{\input{tables/encoders.tex}}
\end{table*}

\begin{figure*}[!t]
\centering

\subfloat[Cross-system FAR Female\label{c1_resnet50}]{
   \includegraphics[width=0.24\linewidth]{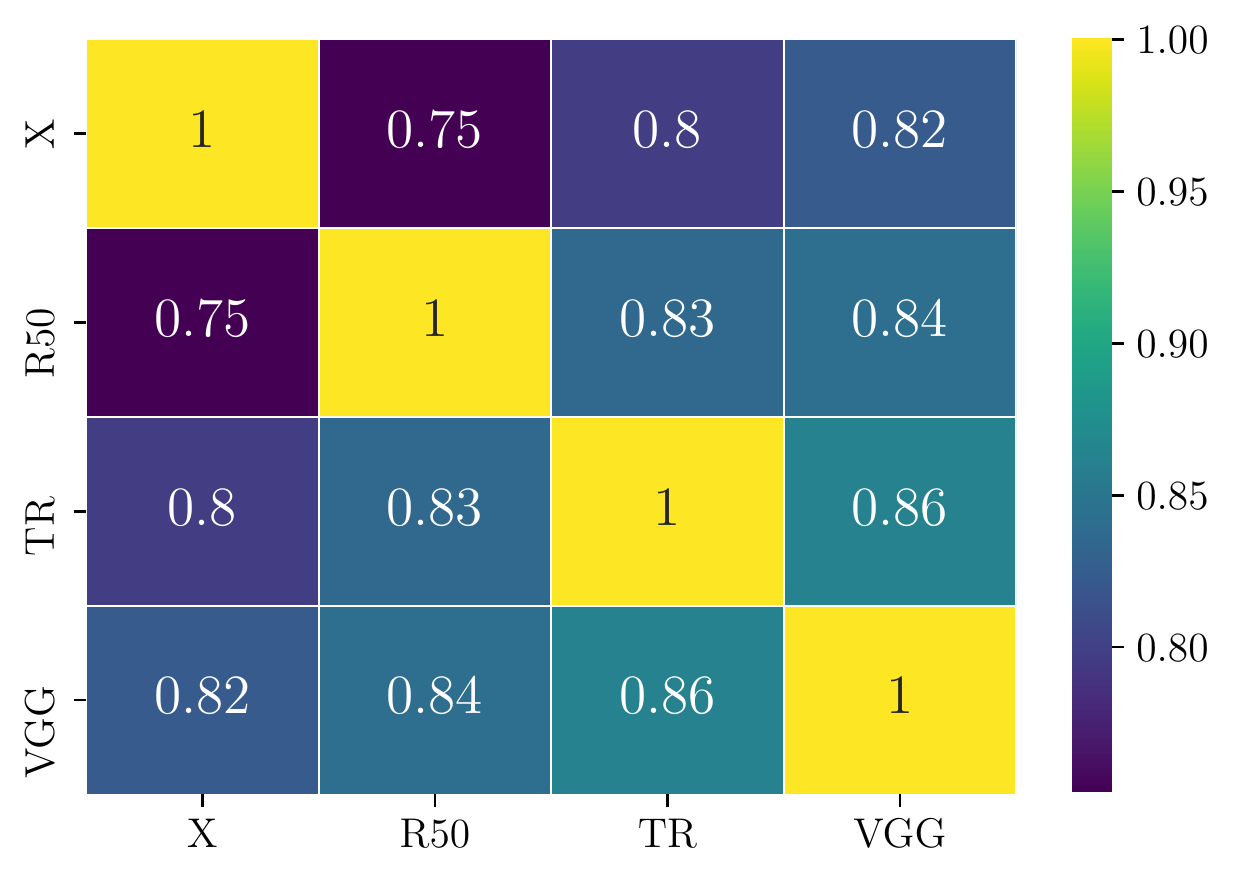}}
\subfloat[Cross-system FAR Male\label{c2_resnet50}]{
   \includegraphics[width=0.24\linewidth]{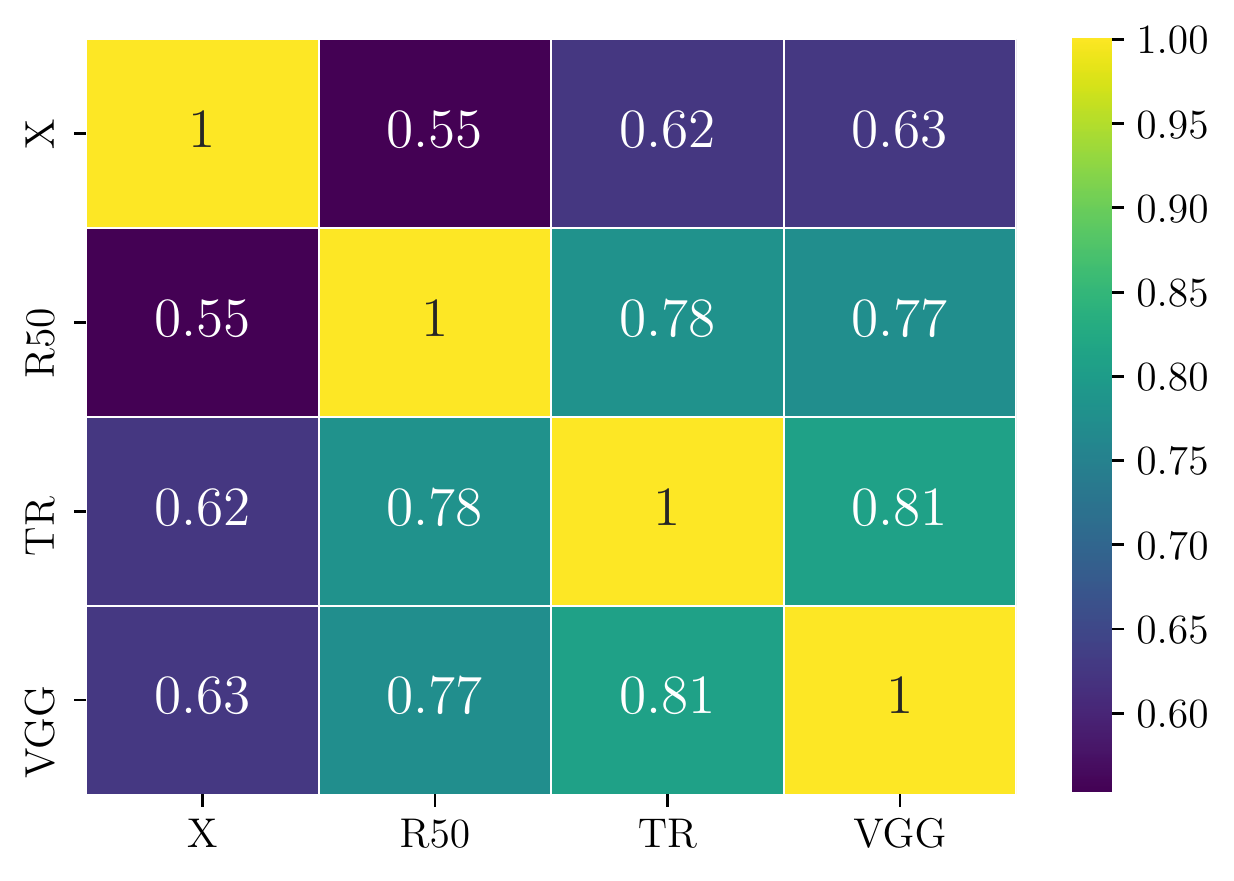}}
\subfloat[Cross-system IR Female\label{c3_resnet50}]{
   \includegraphics[width=0.24\linewidth]{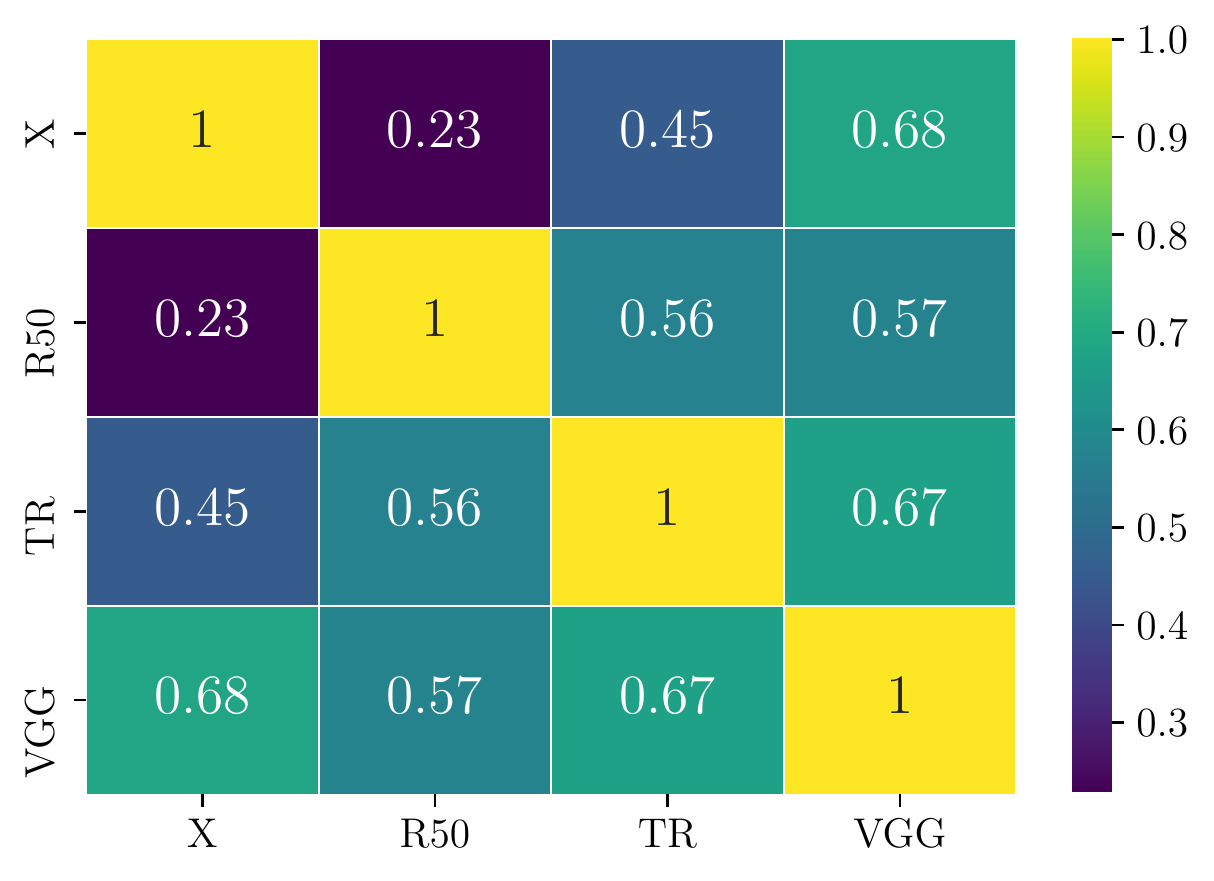}}
\subfloat[Cross-system IR Male\label{c4_resnet50}]{
   \includegraphics[width=0.24\linewidth]{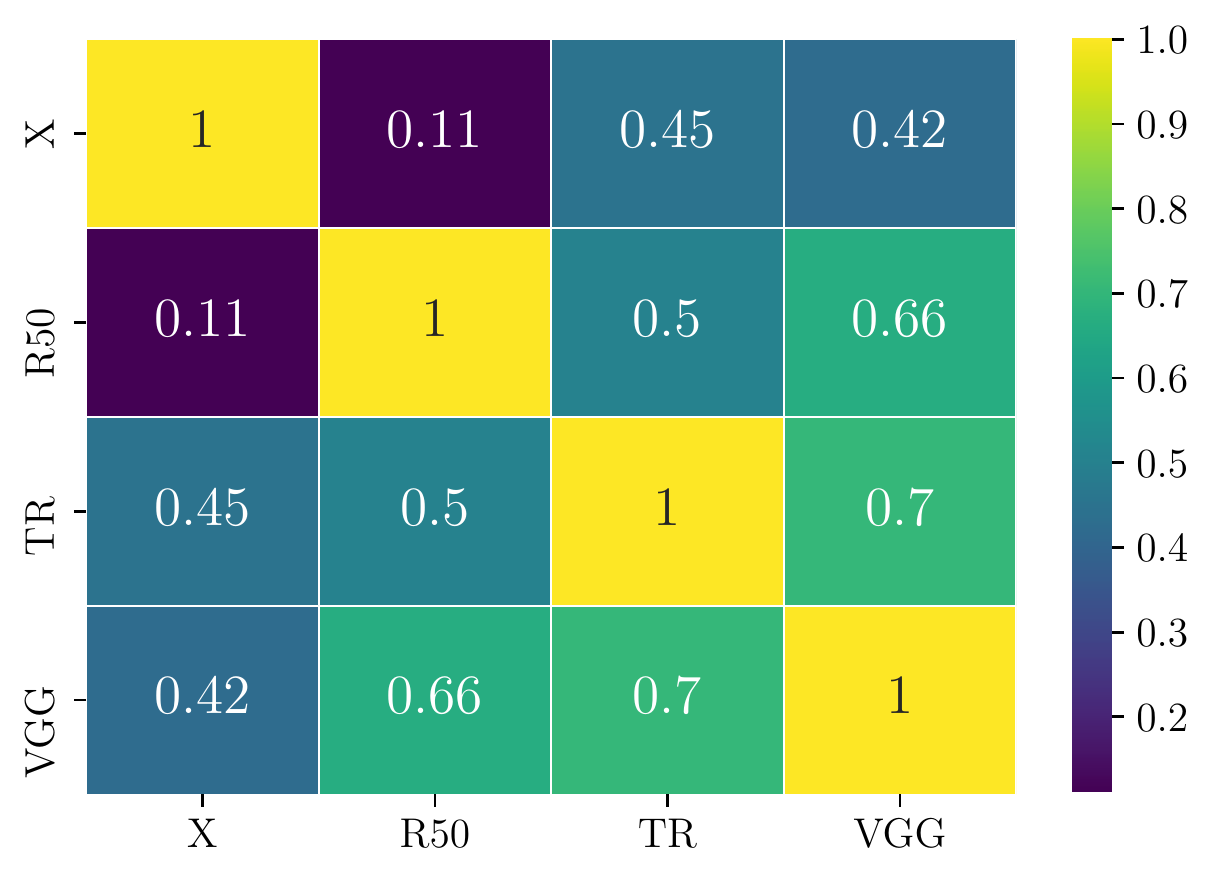}}   
   
\caption{On the left (a) and (b), Spearman correlation respectively between the female and male False Acceptance Rates (FARs) raised by a given seed utterance between two speaker encoders, under an \emph{avg}-10 enrolment and verification policy with a raw \emph{far}-1 threshold. The higher the correlation, the more the utterances tend to have a high impersonation rate on both encoders. On the right (c) and (d), Spearman correlation respectively between the female and male Impersonation Rates (IRs) experienced by the same user between two speaker encoders, under an \emph{avg}-10 enrolment and verification policy with a raw \emph{far}-1 threshold. The higher the correlation, the more the same users end up being impersonated consistently between encoders. It should be noted that FARs and IRs tend to not transfer between CNN-like encoders and x-vector.}
\label{fig:exp-heatmaps}
\vspace{-2mm}
\end{figure*}

\begin{figure*}[!t]
  \includegraphics[width=\textwidth]{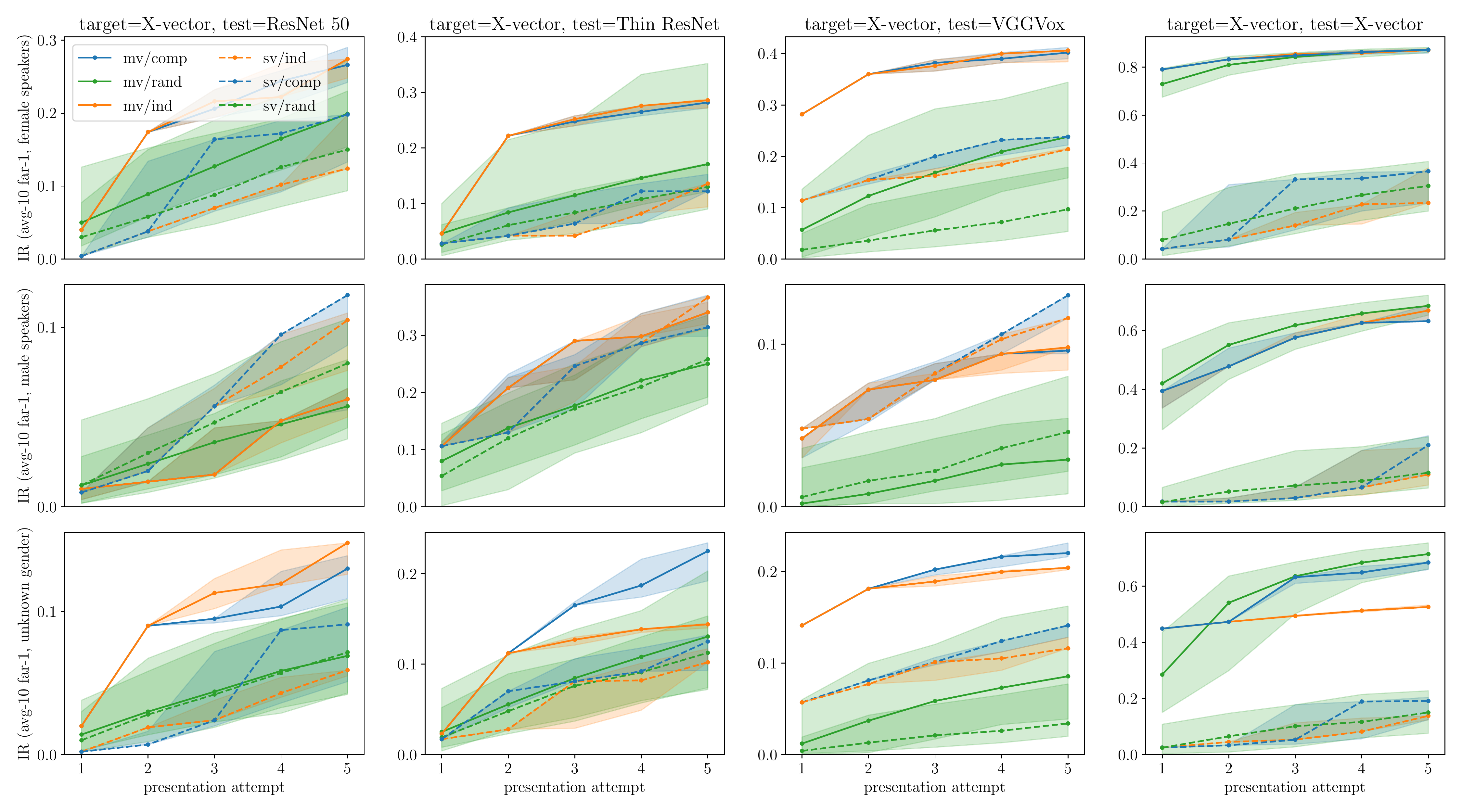}
  \caption{Impersonation rates of seed and master voices under multiple presentation attempts (n=5) in the black-box attack based on NES against the X-vector speaker verification system, under an \emph{avg}-10 enrolment/verification policy with a raw \emph{far}-1 threshold. It should be noted that master voice samples appear to generalize well across populations within X-vector and to transfer well to VGGVox. Transferability performance for the other systems is lower than for the former, though the master voice samples still lead to substantially higher impersonation rates than seed voice examples. }
  \label{fig:black-coveragexv}
\vspace{-2mm}
\end{figure*}

\end{document}

%% file: tables/encoders_compact.tex
\begin{tabular}{lrrrrrrrrrrrr}
\toprule
& \multicolumn{4}{c}{\textbf{AUC}} & \multicolumn{4}{c}{\textbf{EER}} & \multicolumn{4}{c}{\textbf{FRR @ FAR1\%}} \\ 
\cmidrule(lr){2-5} \cmidrule(lr){6-9} \cmidrule(lr){10-13}   
& R$^1$ & R$^2$  & Any & Avg & R$^1$ & R$^2$ & Any & Avg & R$^1$ & R$^2$ & Any & Avg \\
\midrule
\textbf{VGGVox} & 0.95       & 0.98           & 0.90   & 0.93   & 11.8       & 6.9            & 14.5  & 11.3  & 52.1       & 27.0           & 43.2  & 23.2  \\
\textbf{ResNet 50} & 0.96       & 0.98           & 0.92   & 0.94   & 9.9        & 5.2            & 14.0  & 10.8  & 43.7       & 19.9           & 37.6  & 18.6  \\
\textbf{Thin ResNet} & 0.97       & 0.98           & 0.92   & 0.94   &      9.1       & 5.6            & 14.7  & 11.3   &    37.3         &         18.5        & 39.3  & 20.3  \\
\textbf{X-Vector} &     0.96        & 0.97            & 0.91   & 0.93   &       10.9      & 8.2            & 16.0  & 12.5    &       40.2      &       28.2          & 44.8  & 29.2   \\ \bottomrule
\multicolumn{13}{l}{R$^1$ standard VoxCeleb test pairs (no enrollment), tested on short (2.58 s) clips} \tabularnewline
\multicolumn{13}{l}{R$^2$ standard VoxCeleb test pairs (no enrollment), tested on full-length clips}
\end{tabular}

%% file: section_results.tex
\begin{figure*}[t]
  \includegraphics[width=\textwidth]{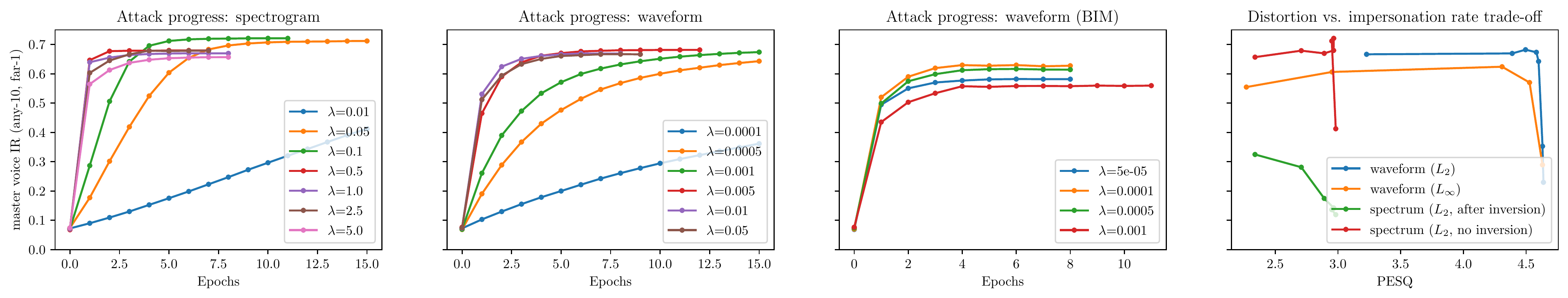}
  \caption{Changes in IR (\emph{any}-10 policy using raw \emph{far-1} threshold) at successive optimization steps: (1st col.) spectrogram optimization; (2nd col.) waveform optimization with updates based on $L_2$-normalized gradient; (3rd col.) waveform optimization with a $L_\infty$ constraint and binarized gradient; trade-off between attack success (impersonation) rate and distortion (PESQ).}
  \label{fig:progress}
  \vspace{-2mm}
\end{figure*}

\section{Evaluation of the Proposed Attack}
\label{sec:results}

In this section, we perform a detailed evaluation of the proposed attack. First, we compare two speech representation domains (waveforms and spectrograms) and investigate the impact of attack and speaker verification settings.
The following experiments focus on a single system configuration (\emph{avg}-$10$, raw \emph{far}-$1$ threshold) and address playback simulation, threat models (white-box vs. black-box) and transferability. We then show efficacy of our attack in a challenging setup with black-box access to a voice cloning system able to generate master voices with arbitrary content. Finally, we consider coverage experiments with multiple presentation attempts.

\subsection{Impact of Attack and Verification Settings}
\label{sec:attack-and-sv-settings}

\begin{figure}
    \includegraphics[width=\columnwidth]{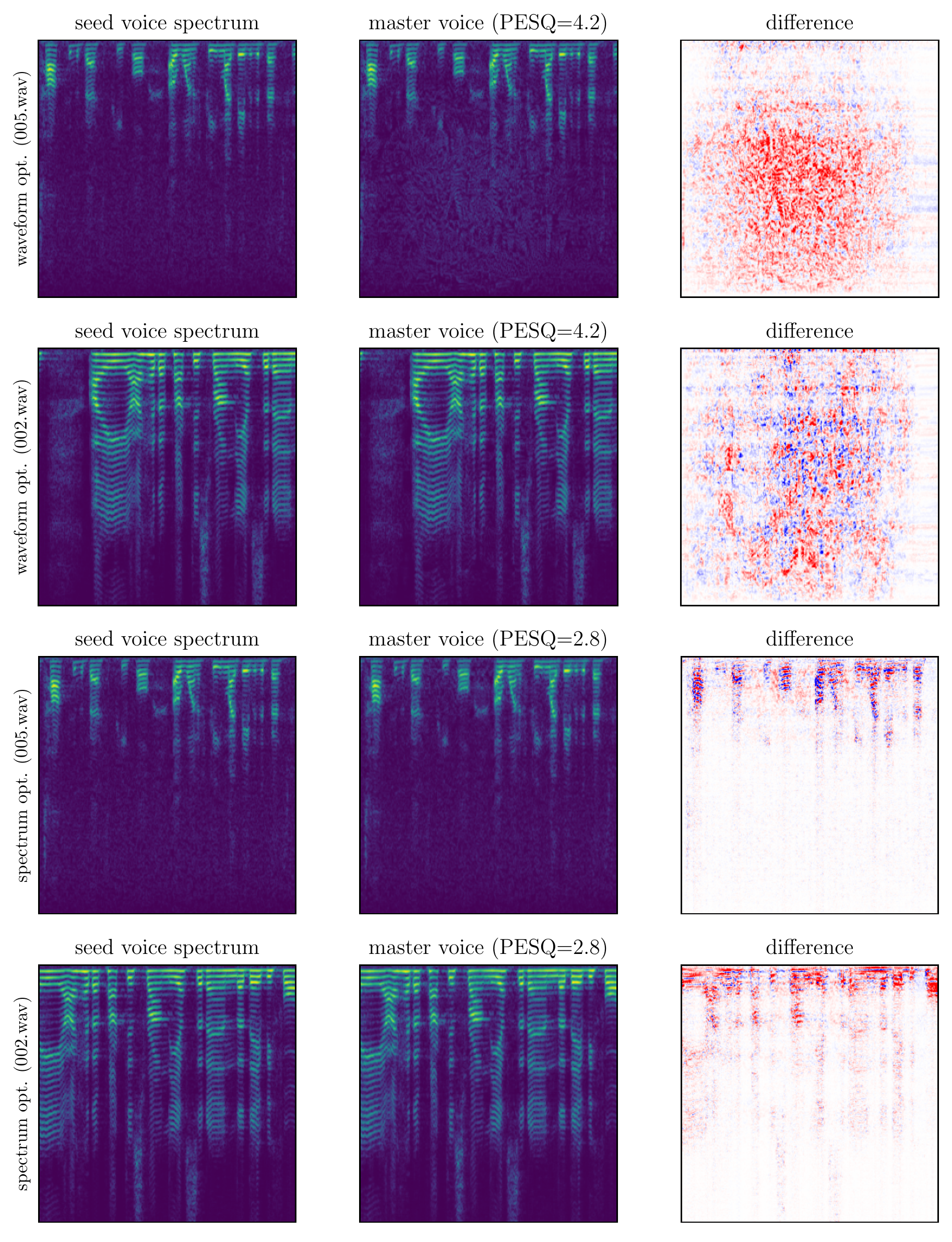}
    \caption{Visualization of frequencies affected by the attack when optimizing the waveform (top 2 rows) and the spectrogram (bottom 2 rows); for the latter, we re-compute the spectrum from a reconstructed adversarial waveform.}
    \label{fig:spectrum-difference}
  \vspace{-2mm}
\end{figure}

\begin{table}[!t]
    \caption{Average IRs for seed voices (SV) and master voices (MV) for various settings scoring strategies.}
    \label{tab:enrollment-policy-breakdown}
    \resizebox{\columnwidth}{!}{\input{tables/a_opt_settings.tex}
}
\end{table}

We first explore the impact of attack and speaker verification settings and measure the success (impersonation) rates. We target the VGGVox encoder and consider various scoring strategies and threshold settings. We consider representative attack variations including optimization in the spectrum and waveform domains and with different update steps, including both $L_2$ and $L_\infty$ normalization and various step sizes $\lambda$ (for the $L_\infty$ variant with budget $\epsilon$, we use steps size $\lambda=\frac{\epsilon}{10}$ to allow for more flexibility).

Fig.~\ref{fig:progress} shows how the female IR changes with successive epochs (passes over the entire population) for various step sizes ($\lambda$). Successive columns correspond to spectrum optimization with $L_2$ gradient normalization (1st column), and waveform optimization with $L_2$ (2nd) and $L_\infty$ normalization (3rd). We can observe that our attack is highly effective - it substantially improves IRs across various settings and transfers well between user populations. For the monitored \emph{any}-$10$ policy, the average impersonation rate on the unseen population increases from 7\% to $\approx$66\%. Convergence rates vary with step size, but the attack saturates at a comparable level. $L_\infty$ tends to converge faster than $L_2$, but reaches lower success rates across all distortion levels (see column 4). 

While at the time of the attack waveform and spectrogram optimization seem to reach similar IRs, the latter requires spectrogram inversion to yield an adversarial waveform (we used the Griffin-Lim algorithm). While the attack still works, it operates at an evidently reduced efficacy (down to $\approx20-30\%$ in this experiment) and suffers from reduction of audio fidelity (see the gap in Perceptual Evaluation of Speech Quality PESQ~\cite{rix2001perceptual} scores in column 4; higher scores for higher audio fidelity). We also visually compare the character of adversarial distortions in Fig.~\ref{fig:spectrum-difference}. Top rows depict 2 pairs of seed-master voice samples got with waveform optimization, and bottom rows depict 2 pairs of seed-master voice samples got with spectrum optimization. Waveform optimization affects a wider and higher range of frequencies.

In Table~\ref{tab:enrollment-policy-breakdown}, we summarize IRs obtained for both female and male populations across several enrollment policies and decision thresholds. For clarity, we report only one set of master voice samples which corresponds to a good trade-off between efficacy and audio fidelity. Our attack is effective regardless of system configuration and achieves non-trivial matching rates even in the most restrictive setting. At a \emph{far}-$1$ threshold calibrated for the \emph{avg}-10 policy, our master voice samples still impersonate 69\% of females and 38\% of males in a population unknown to the attacker.

Based on these results, in subsequent experiments we will restrict our attention to waveform-based attacks with $L_2$ gradient normalization and to speaker verification based on the \emph{avg}-$10$ policy operating at a raw \emph{far}-$1$ threshold.

\subsection{Experiments with Playback Simulation}

\begin{figure}
    \includegraphics[width=\columnwidth]{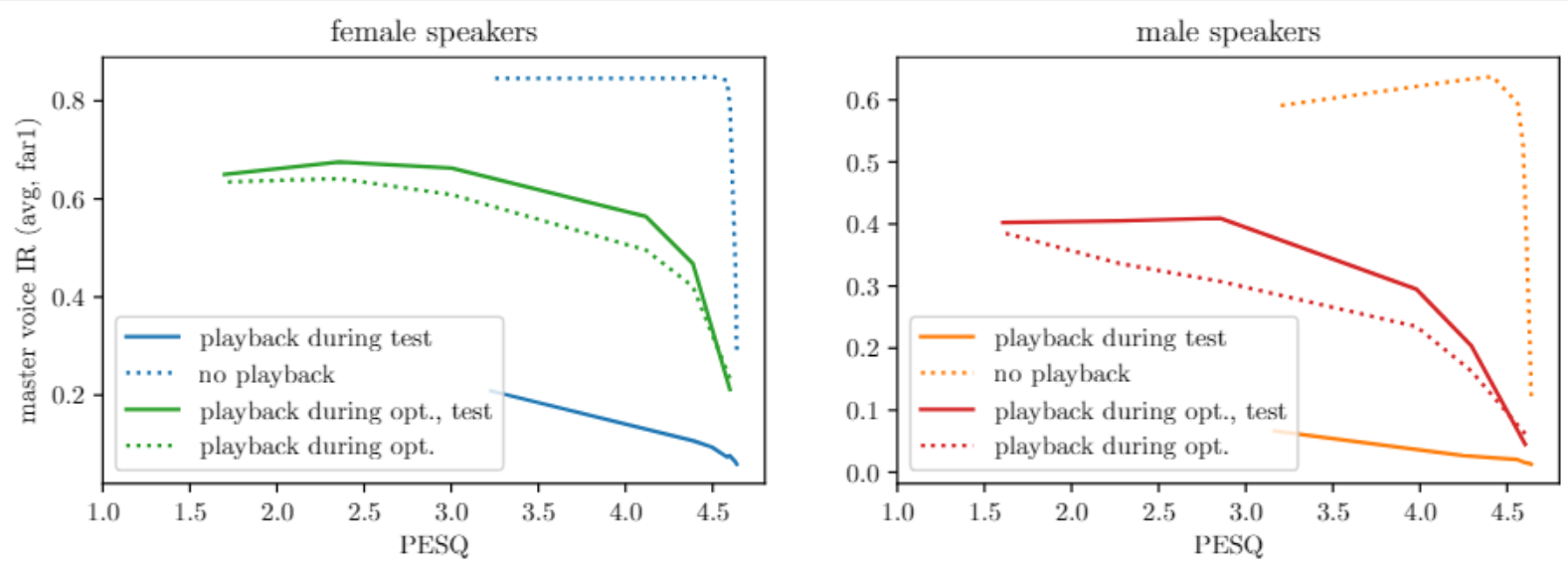} \\
    \includegraphics[width=\columnwidth]{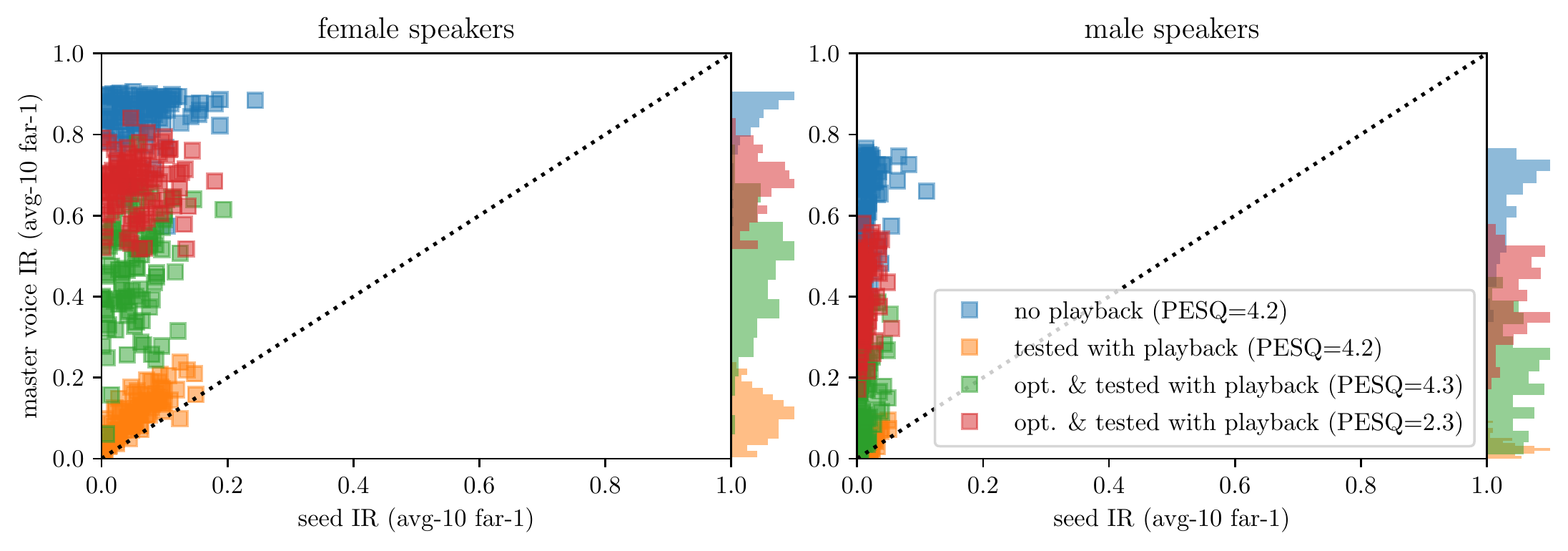}
    \caption{Impact of playback simulation at the time of testing and optimization: (top) trade-off between IRs; (bottom) detailed scatter plot of (seed, master) IRs for selected configurations.}
    \label{fig:playback}
    \vspace{-2mm}
\end{figure}

\begin{figure}
  \includegraphics[width=\columnwidth]{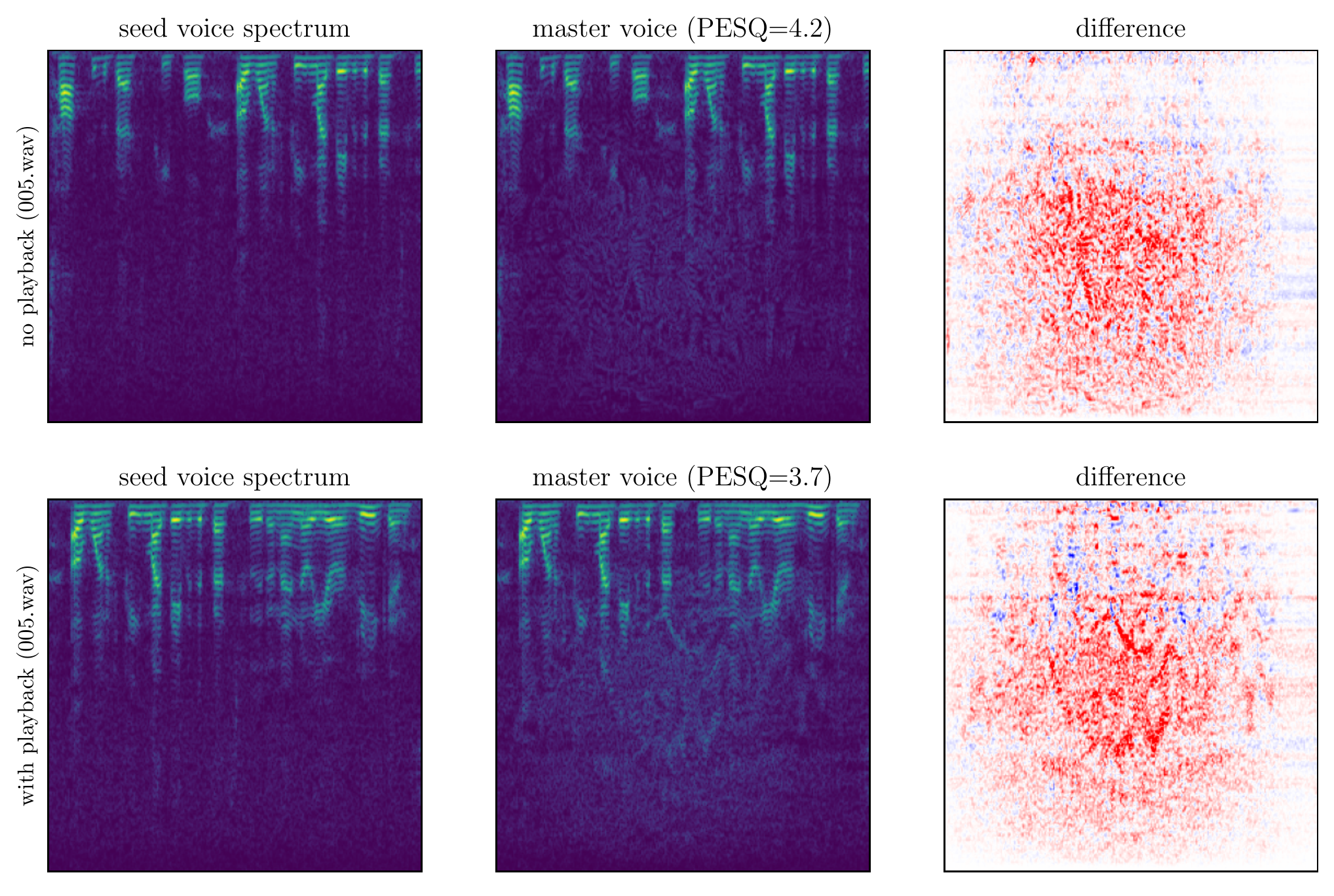}
  \caption{Visualization of frequencies affected by waveform optimization with and without playback simulation.}
  \label{fig:playback-spectrum}
    \vspace{-2mm}
\end{figure}

To test robustness of our attack to various distortions, we implemented playback simulation, which combines additive Gaussian noise with characteristics of a speaker, microphone, and surrounding environment (Section~\ref{sec:playback}). The simulation can be included both at testing and optimization time - assuming the representation domain precedes playback (e.g,. waveform or speaker embedding in synthesis systems). We therefore experiment with waveform optimization and assess the impact of the distortion at each of the mentioned stages. 

In the following description, we use the terms \emph{standard} and \emph{augmented} optimization to indicate presence of playback. We randomly choose playback settings (noise strength, impulse response) for each batch. Apart from this, we follow the same experimental setup as before, i.e., we vary step size $\lambda$ and measure IRs at various distortion levels. We show the obtained results in Fig.~\ref{fig:playback}. The top row illustrates the trade-off between master voice IR and speech fidelity (PESQ). We can observe that without augmented optimization, test-time playback (solid lines) renders adversarial waveforms nearly ineffective. The attack success rate can still increase somewhat, despite obvious saturation (or even rebound), during a standard test.

Augmentation leads to much more robust adversarial examples that achieve similar success rates with(out) playback (see solid vs. dotted lines of the same color). It leads to larger distortion, but does not obviously alter which frequencies are affected (Fig.~\ref{fig:playback-spectrum}). In both cases, the distortion tends to be the strongest in the middle of the sample. When audible, it sounds like a hissing modulated noise that does not interfere with the spoken content or the perceived identity of the speaker.

\begin{figure}[!t]
  \includegraphics[width=\columnwidth]{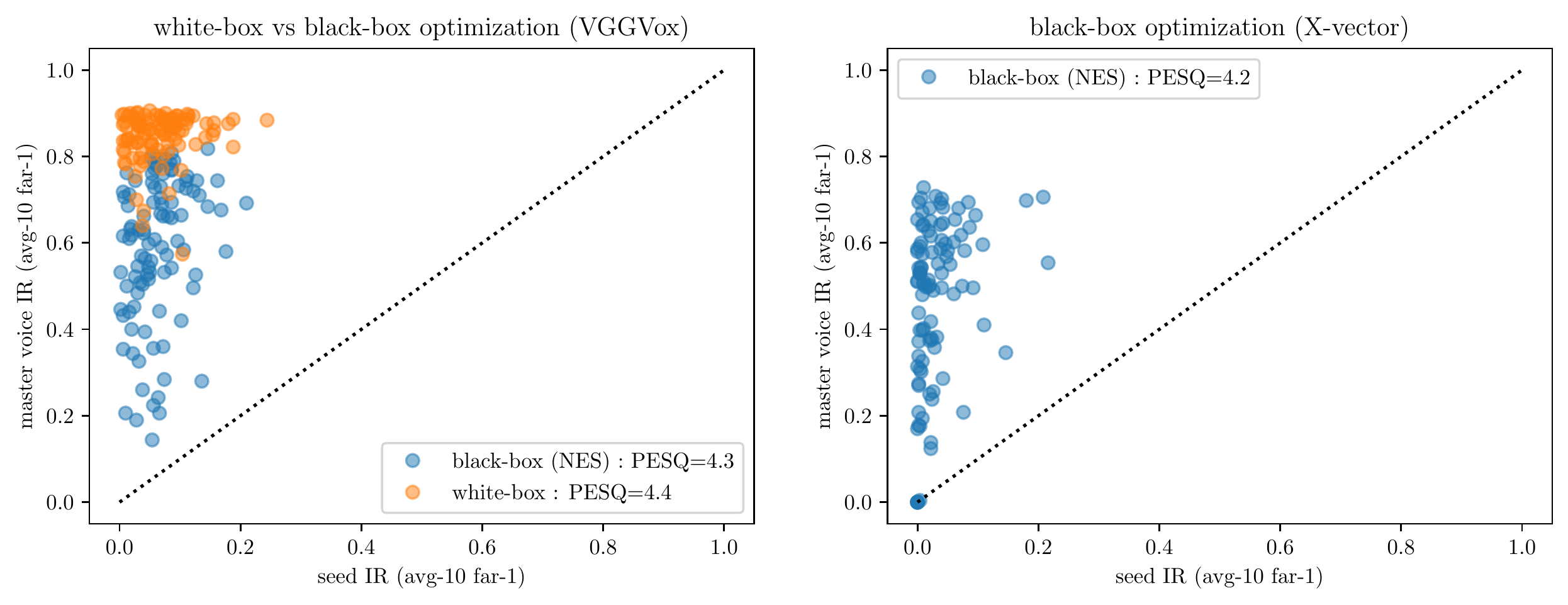}
  \caption{Scatter plots of seed-master IRs for white-box optimization with accurate gradients vs. black-box optimization with NES-estimated gradients: (left) results for VGGVox at a similar distortion level; (right) results for x-vector where white-box optimization was not possible.}
  \label{fig:black-box}
  \vspace{-2mm}
\end{figure}

\begin{figure}[!t]
  \includegraphics[trim=0 200 0 0,clip,width=\columnwidth]{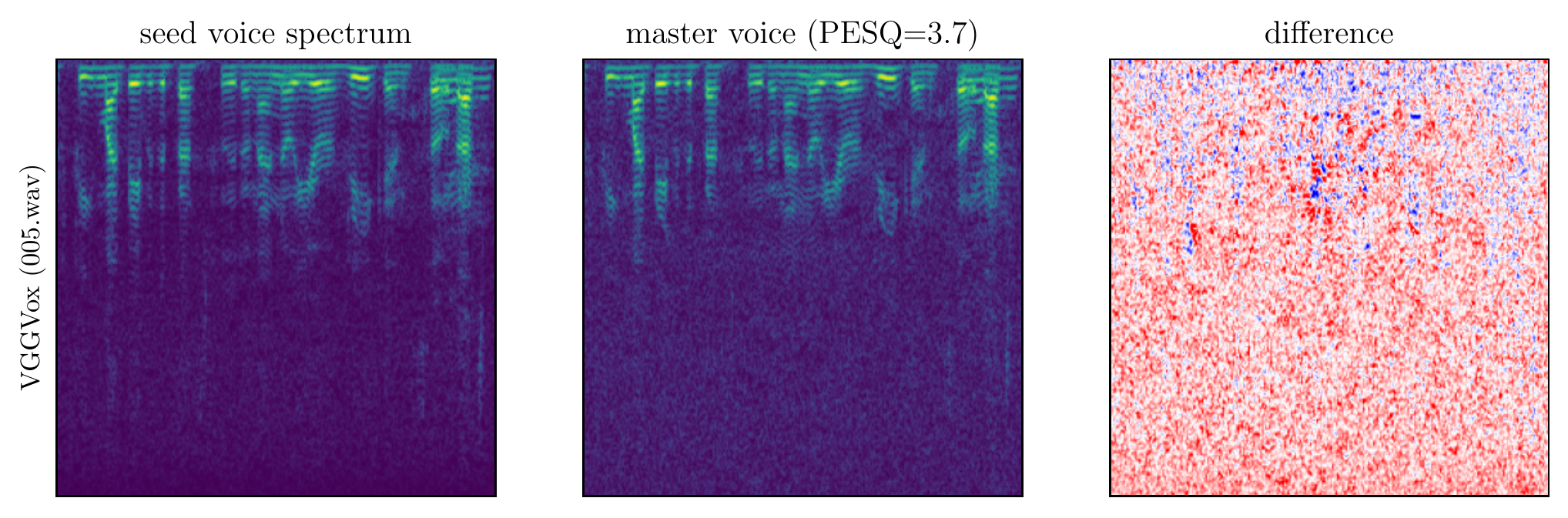}
  \caption{Visualization of frequencies affected by waveform optimization in the black-box attack based on NES.}
  \label{fig:black-box-spectrum}
  \vspace{-2mm}
\end{figure}

\subsection{White-box vs Black-box Attacks}
\label{sec:black-box-evaluation}

\begin{table}[!t]
    \caption{Transferability of master voice obtained with waveform optimization targeting different speaker encoders}
    \label{tab:transferability}
    \resizebox{\columnwidth}{!}{\input{tables/d_transferability_new.tex}}
\end{table}

Previous experiments relied on full gradients provided by automatic differentiation features in Tensorflow. While this leads to an effective attack, it is inflexible and often even impractical - due to either lack of knowledge or excessive implementation time to make everything fully differentiable. We hence switch to black-box optimization with gradients estimated by NES (Section~\ref{sec:black-box-intro}) from a similarity score returned by the speaker verification system. Based on preliminary experiments on a small grid of feasible values, we set NES parameters to $s=100$ samples and $\sigma=0.001$. Due to large increase in computational requirements, we use a single step size $\lambda=0.01$ and limit the number of epochs\footnote{NES relies on many independent function calls, trivially parallel, and could be optimized. Our naive sequential implementation on one RTX 8000 GPU required approx. 10 minutes per epoch (with $s=100$ function samples). The corresponding white-box optimization takes approx. 10 seconds per epoch.} to 10.

Fig.~\ref{fig:black-box} compares white-box and black-box optimization for VGGVox (left) and shows black-box results for x-vector which uses non-differentiable filter-banks as its acoustic representation (right). In both cases, our black-box attack reaches IRs of 47\% for x-vector and 59\% for VGGVox (white-box attack at a comparable distortion reached 85\%). Compared to white-box optimization, the black-box attack uniformly affected all frequencies at all times (Fig.~\ref{fig:black-box-spectrum}). 

\subsection{Experiments with Transferability}

We then test master voice transferability between speaker encoders: ResNet 50, Thin ResNet, VGGVox (all based on spectrograms) and x-vector (based on filter banks). We used waveform optimization in the white-box setting and fall back to black-box NES updates for x-vector. We test all combinations of playback simulation (optimization and testing time).

We collected results in Table~\ref{tab:transferability} (female speakers). In general, waveform optimization does not lead to transferable master voices. The obtained adversarial speech relies on carefully crafted noise (see Fig.~\ref{fig:spectrum-difference}, \ref{fig:playback-spectrum}, \ref{fig:black-box-spectrum}) and not on changes in speaker characteristics. Playback simulation in augmented training does provide a small but consistent improvement in transferability, but insufficient for an effective attack.

\begin{figure*}[!t]
  \includegraphics[width=\textwidth]{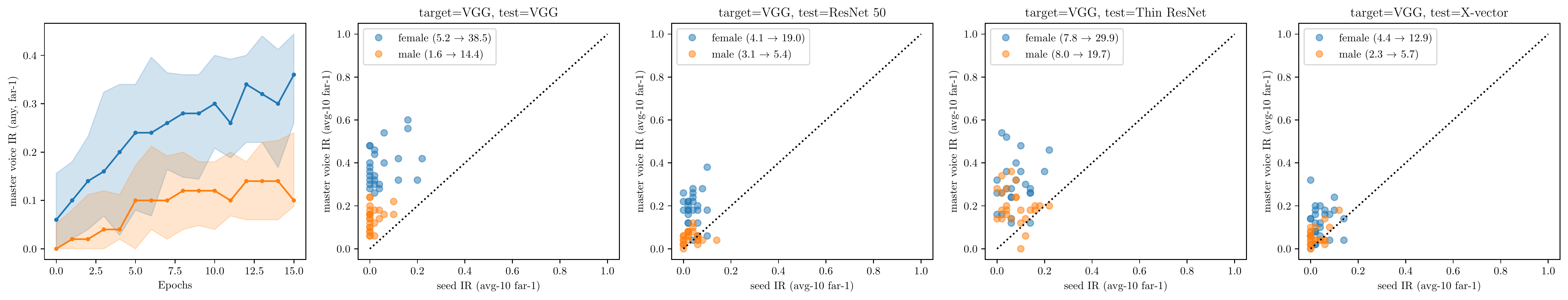} \\
  \includegraphics[width=\textwidth]{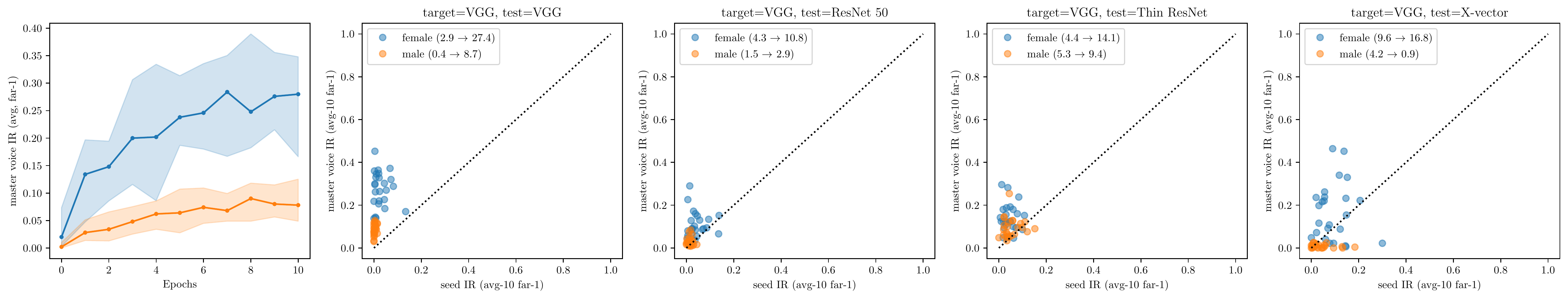}
  \includegraphics[width=\textwidth]{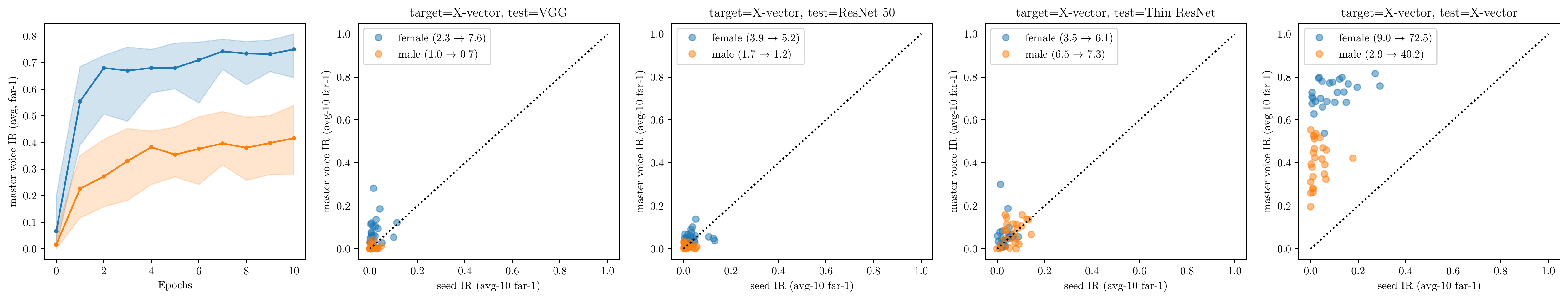}
  \vspace{-8mm}
  \caption{Optimization progress and final impersonation rates of seed-master voice pairs obtained with voice cloning: (top) optimization targeting the VGGVox encoder on the LibriSpeech dataset; (middle) optimization targeting VGGVox on the VoxCeleb dataset; (bottom) optimization targeting x-vector on the VoxCeleb dataset. Targeting VGGVox tends to transfer to other models with best results observed for female speakers and other spectrogram-based encoders (ResNets). Targeting x-vector yielded stronger impersonation capabilities but poor transferability.}
  \label{fig:cloning}
  \vspace{-2mm}
\end{figure*}

\subsection{Experiments with Voice Cloning}
\label{sec:results-cloning}

Optimization in the waveform domain leads to highly effective adversarial speech samples (reaching even up to 85\% IR) that can be made robust to various distortions via playback simulation. However, the optimization learns to embed carefully crafted noise that does not change the content or speaker identity and generally does not transfer between encoder architectures. In this section, we take advantage of the flexibility of our attack and experiment with a more compact, disentangled representation - we investigate optimization of the speaker embedding in a complex voice cloning system. 

We used an open source system~\cite{rtvc,jia2018transfer} that generates speech based on a text prompt and a 256-d speaker embedding. The system uses Tacotron~\cite{DBLP:conf/interspeech/WangSSWWJYXCBLA17} for waveform synthesis, WaveRNN~\cite{DBLP:conf/icml/KalchbrennerESN18} as a vocoder, and an LSTM-based encoder~\cite{DBLP:conf/icassp/WanWPL18}. All models are implemented in PyTorch, and we integrated the system with our Tensorflow-based attack framework via a simple black-box API that exposes two functions:

\begin{footnotesize}
\begin{verbatim}
- get_speaker_embedding(speech_sample)
- generate_speech(text, speaker_embedding, max_len)
\end{verbatim}
\end{footnotesize}

\noindent The generated output is stochastic and exhibits variations in sound and length of the waveforms. As a result, it represents a realistic and challenging attack scenario.

We fixed the text prompt to \emph{"The assistant is triggered by saying hey google"} and use NES to evolve the speaker embedding, initialized from a seed voice by the black box speaker encoder $\mathcal{E}'$. Based on preliminary experiments, we set NES parameters to $s=50$, $\sigma=0.025$ and step size to $\lambda=0.1$. We also clip\footnote{
\color{black}
Clipping was performed to avoid deviations from the domain of seed vectors of the generative model. Departure from the commonly used n-dimensional hypercube or high-density regions of a standard multivariate Gaussian tend to introduce artifacts or break the synthesis entirely.} the embedding to stay within the expected domain of $[0, 1]^{256}$ and normalize the length of the output waveforms to 2.58 seconds. The cloning system was trained using LibriSpeech~\cite{panayotov2015librispeech}, so we conduct our experiments on this dataset as well. On VoxCeleb, we used the same setup as before. On LibriSpeech, we used a popular subset \emph{train-clean-100} with 250 speakers which we split into two disjoint populations (optimization and testing) with 100 people, each with equal balance between genders. We randomly chose 10 samples per speaker both for optimization and for enrollment into the verification system. Due to much larger computational footprint, we repeat the attack based on 25 seed samples and run the attack for 15 (or 10) epochs for the LibriSpeech (VoxCeleb) datasets, respectively. We target the VGGVox and x-vector encoders and assess transferability.

\begin{figure}[!t]
    \includegraphics[width=\columnwidth]{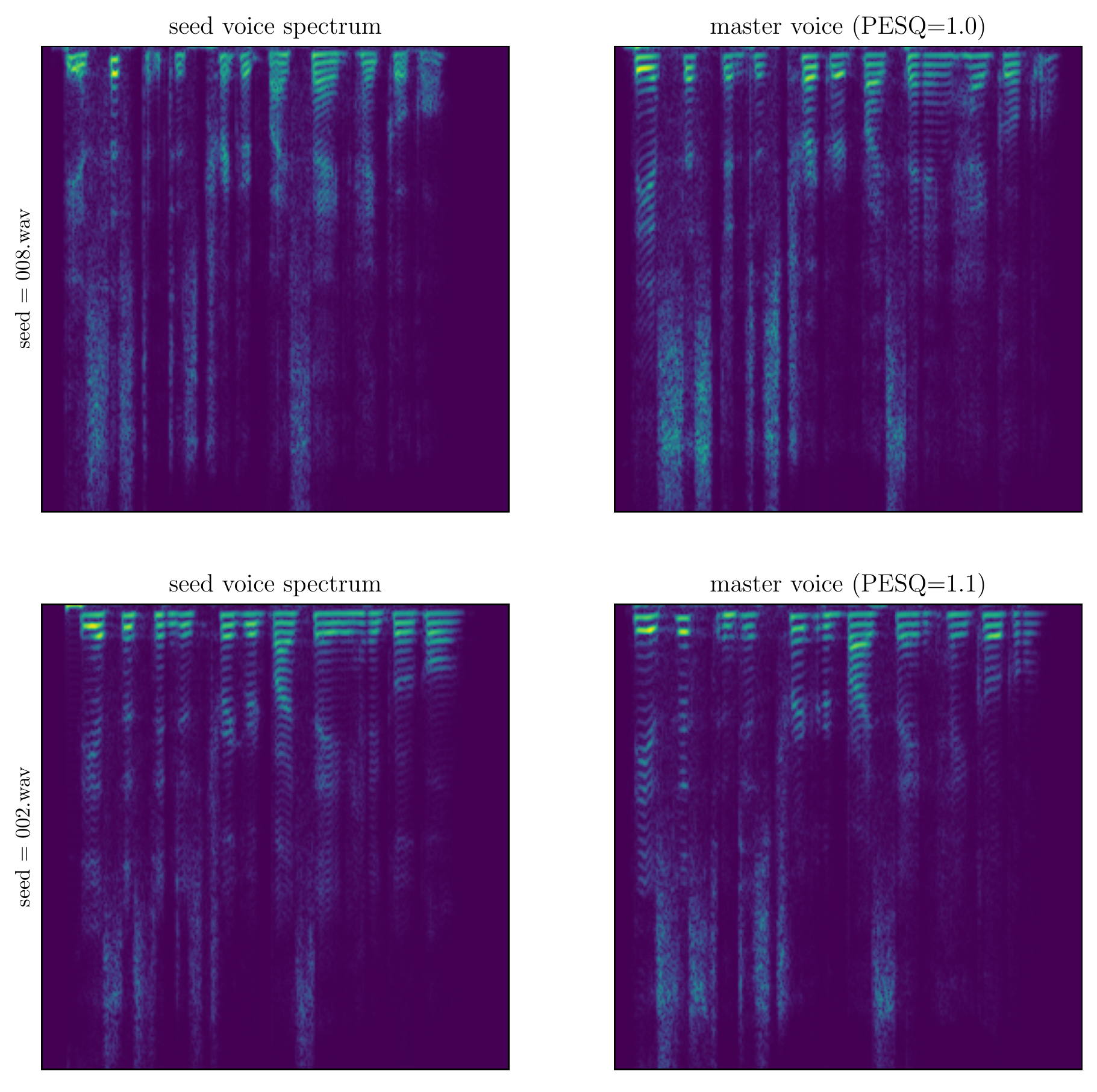}
    \caption{Spectrograms of two pairs of seed and master voice samples obtained with voice cloning; the attack adapts the speaker characteristics and does not in result in tailored adversarial noise.}
    \label{fig:cloning_spectrograms}
  \vspace{-2mm}
\end{figure}

\begin{table}
  \caption{Transferability of master voices obtained with voice-cloning}
  \label{tab:cloning}
  \resizebox{\columnwidth}{!}{\input{tables/f_cloning.tex}}
\end{table}

We show the obtained results in Fig.~\ref{fig:cloning}. The left column shows attack progress for male and female speakers along with top and bottom percentiles (90-th and 10-th, respectively) of the observed impersonation rates (on the unseen test population). Despite the randomness of the generation and large variations in numbers, our attack consistently increases IRs for both male and female speakers, although the effect is substantially stronger for the latter. We compare the initial and final IRs using scatter plots (columns 2-5) for all considered speaker encoders. Each row corresponds to one targeted model (VGGVox or x-vector) and successive columns correspond to different test models and demonstrate transferability of the obtained samples. On the small LibriSpeech dataset (1st row) master voices optimized using VGGVox successfully transferred across all encoders. Again, the effect depends on the gender and tends to be much stronger for female speakers. On VoxCeleb the results are similar with the exception of male speakers tested on x-vector (which surprisingly has a strong negative effect). Targeting x-vector yielded much more effective, but generally non-transferable master voice samples - although for female speakers a weak effect seems to exist. We summarize the average impersonation rates in Table~\ref{tab:cloning}.

In contrast to waveform optimization, the attack does not result in obvious adversarial artifacts and appears to adapt the speaker characteristics (see  Fig.~\ref{fig:cloning_spectrograms}). This may explain improved transferability between speaker encoders (Table~\ref{tab:cloning} and Fig.~\ref{fig:cloning}) and appears to match cross-system biometric menagerie evaluation. We assessed correlations of impersonation rates between all systems as a further validation (see Fig.~\ref{fig:exp-heatmaps}).

\begin{figure*}
  \includegraphics[width=\textwidth]{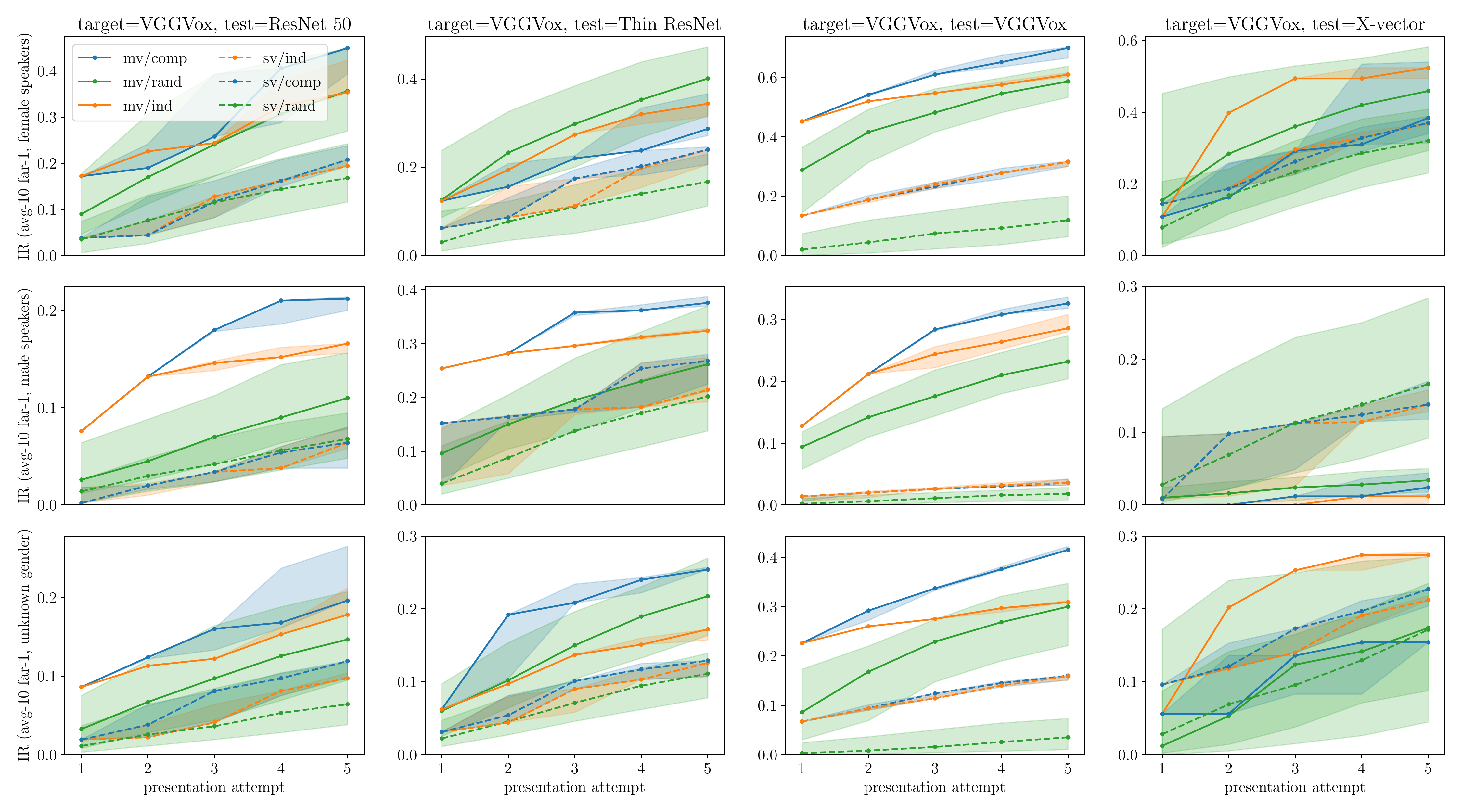}
  \vspace{-8mm}
  \caption{Impersonation rates of seed and master voices under multiple presentation attempts ($n=5$) in the black-box attack based on NES against the VGGVox speaker encoder, under an \emph{avg}-10 policy with raw \textit{far-1} threshold. (top) female gender, (middle) male gender, (bottom) unknown gender.}
  \label{fig:black-coverage}
  \vspace{-2mm}
\end{figure*}

\subsection{Experiments with Multiple Presentation Attempts}
\label{sec:results-coverage}

We finally evaluated seed and master voices in a setting where the speaker verification system allows users to do more than one attempt (we considered $c=5$ allowed attempts). We tested two simple strategies (see Section \ref{sec:multi-presentation}): naive independent selection (\emph{ind}) and complementary selection (\emph{comp}). To obtain more stable results, a strategy was repeated 100 times, each on a different subset of the seed/master voice population composed by 75\% of randomly sampled users (results were averaged). We compared the two strategies against a random selection (\emph{rand}). Due to their high impersonation power and transferability, we focus on master voice examples optimized (with playback) for the speaker embedding in a voice cloning system, targeting VGGVox (see Section \ref{sec:results-cloning}).

We collected the results for VGGVox in Fig.~\ref{fig:black-coverage}, for each gender separately (first two rows) and for a setting where we assume the attacker does not know the victim's gender (third row). In general, a complementary selection on master voices leads to the highest overall and cross-attempt IRs, except for x-vector. The adversarial speech relies on carefully crafted perturbations targeting a CNN-like architecture (VGGVox), and those perturbations might not be comparably effective on a different type of encoder architecture (x-vector is a TDNN based on filter banks; the others are CNN based on spectrograms). This is confirmed also by a transferability analysis in Fig. \ref{fig:exp-heatmaps}, which shows that FARs and IRs for seed utterances tend to not transfer between CNN-like encoders and x-vector.
The explored selections obtain substantial gains for the male speakers, often doubling the IR of the best seed setting at that attempt. These two strategies also allow to improve transferability against Thin ResNet and ResNet 50. Similar observations were made while targeting x-vector during optimization (Fig. \ref{fig:black-coveragexv}).

%% file: tables/a_opt_settings.tex

\begin{tabular}{lrrrrm{0em}rrrr}
\toprule
 & \multicolumn{4}{c}{\textbf{waveform optimization}} &  & \multicolumn{4}{c}{\textbf{spectrum optimization}}\tabularnewline
\cmidrule(lr){2-5} \cmidrule(lr){7-10}
 & \multicolumn{2}{c}{\textbf{female}} & \multicolumn{2}{c}{\textbf{male}} &  & \multicolumn{2}{c}{\textbf{female}} & \multicolumn{2}{c}{\textbf{male}}\tabularnewline
 \cmidrule(lr){2-3} \cmidrule(lr){4-5} \cmidrule(lr){7-8} \cmidrule(lr){9-10}
 & \textbf{SV} & \textbf{MV} & \textbf{SV} & \textbf{MV} &  & \textbf{SV} & \textbf{MV} & \textbf{SV} & \textbf{MV}\tabularnewline

\midrule
any, $\tau :$ far1$^1$ &  7.3 & 66.9 &  2.1 & 21.2 &  & 7.3 & 67.9  &  2.2  & 23.2 \tabularnewline
any, $\tau :$ far1     &  7.3 & 67.0 &  2.0 & 21.2 &  & 7.2 & 28.1  &  2.1  &  5.7 \tabularnewline
any, $\tau :$ eer      & 37.5 & 96.1 & 17.1 & 91.7 &  & 37.6 & 74.5 & 17.7 & 39.0 \tabularnewline
\midrule
avg, $\tau :$ far1     &  6.9 & 84.7 &  1.6 & 63.3 &  & 6.7 & 37.7  &  1.7  & 10.7 \tabularnewline
avg, $\tau :$ far1$^2$ &  2.4 & 69.5 &  0.4 & 38.0 &  & 2.5 & 19.4  &  0.5  &  3.5 \tabularnewline
avg, $\tau :$ eer      & 32.3 & 96.7 & 13.0 & 97.1 &  & 32.0 & 74.7 & 13.9 & 39.8 \tabularnewline
\bottomrule
\multicolumn{10}{l}{$^1$ optimization-time measurements without spectrogram inversion} \tabularnewline
\multicolumn{10}{l}{$^2$ attack with good performance at a low distortion level}
\end{tabular}

%% file: tables/d_transferability_new.tex
\begin{tabular}{lrrrrm{0em}rrrr}
\toprule
\multirow{2}{*}{\diagbox{\textbf{Target}$^2$}{\textbf{Test}$^1$}} & \multicolumn{4}{c}{\textbf{tested w/o playback}} &  & \multicolumn{4}{c}{\textbf{tested w/ playback}}\tabularnewline
\cmidrule(lr){2-5} \cmidrule(lr){7-10}
    & \textbf{R50} & \textbf{TR} & \textbf{VGG} & \textbf{X} &  & \textbf{R50} & \textbf{TR} & \textbf{VGG} & \textbf{X}\tabularnewline
\midrule
& \multicolumn{9}{c}{\textbf{standard optimization}} \tabularnewline
\midrule
MV : ResNet 50  &  \textbf{35.7} & 4.4 & 4.5 & 4.1 & & \textbf{4.8} & 2.5 & 3.7 & 0.4 \tabularnewline
MV : Thin ResNet & 3.2  & \textbf{68.6} & 5.2 & 4.4 & & 2.8 & \textbf{7.1} & 4.1 & 0.4  \tabularnewline
MV : VGG & 2.7 & 5.5 & \textbf{89.6} & 4.7 & & 2.9 & 2.6 & \textbf{9.3} & 0.4 \tabularnewline
MV : X-vector$^3$ &   5.1 & 9.0 & 7.4 & \textbf{73.3} & & 5.3 & 4.8 &  5.3 & \textbf{1.6} \tabularnewline
\midrule
SV : seed voice & 2.5  & 4.6 & 4.7 & 3.5 & &  2.8 & 2.4 & 3.8 & 0.5 \tabularnewline
\midrule
& \multicolumn{9}{c}{\textbf{augmented optimization}} \tabularnewline
\midrule
MV : ResNet 50 & \textbf{26.8} & 4.9 & 6.1 & 4.2 & & \textbf{33.0} & 3.0 & 5.3 & \textbf{0.4} \tabularnewline
MV : Thin ResNet & 3.3   & \textbf{43.3} & 7.3  & \textbf{4.9} & & 3.3 & \textbf{51.9} & 6.5 & 0.4 \tabularnewline
MV : VGG &  3.4  &  6.3 & \textbf{36.1} & 5.1 & &  3.4& 3.5  & \textbf{39.2} & 0.5 \tabularnewline
\midrule
SV : seed voice &  2.5 & 4.5 & 4.8 & 3.6 & & 2.9 & 2.3 & 4.1 & 0.4 \tabularnewline
\bottomrule
\multicolumn{10}{l}{$^1$ Abbrev.: (R50) ResNet 50; (TR) Thin ResNet; (VGG) VGGVox; (X) X-Vector} \tabularnewline
\multicolumn{10}{l}{$^2$ Master voice examples were optimized with $\lambda=0.01$} \tabularnewline
\multicolumn{10}{l}{$^3$ uses black-box optimization} 
\end{tabular}

%% file: tables/f_cloning.tex
\begin{tabular}{lccccm{0em}cccc}
    \toprule
     & \multicolumn{4}{c}{\textbf{seed voice}} &  & \multicolumn{4}{c}{\textbf{master voice}}\tabularnewline
    \cmidrule(lr){2-5} \cmidrule(lr){7-10}     
     \cmidrule(lr){2-3} \cmidrule(lr){4-5} \cmidrule(lr){7-8} \cmidrule(lr){9-10}
     & \textbf{R50} & \textbf{TR} & \textbf{VGG} & \textbf{X} &  & \textbf{R50} & \textbf{TR} & \textbf{VGG} & \textbf{X}\tabularnewline    
    \midrule
    & \multicolumn{9}{c}{\textbf{(LibriSpeech) far-1 calibrated on raw embedding similarity}} \tabularnewline
    \midrule
    VGG (female)    & 4.1 & 7.8 & 5.2 & 4.4 & & 19.0 & 29.9 & \textbf{38.5} & 12.9 \tabularnewline
    VGG (male)      & 3.1 & 8.0 & 1.6 & 2.3 & & 5.4 & 19.7 & \textbf{14.4} & 5.7 \tabularnewline
    \midrule
    & \multicolumn{9}{c}{\textbf{(VoxCeleb) far-1 calibrated on raw embedding similarity}} \tabularnewline
    \midrule
    VGG (female)    & 4.3 & 4.4 & 2.9 & 9.6 & & 10.8 & 14.1 & \textbf{27.4} & 16.8 \tabularnewline
    VGG (male)      & 1.5 & 5.3 & 0.4 & 4.2 & & 2.9 & 9.4 & \textbf{8.7} & 0.9 \tabularnewline
    \midrule
    & \multicolumn{9}{c}{\textbf{(VoxCeleb) far-1 calibrated on raw embedding similarity}} \tabularnewline
    \midrule
    X-vector (female)    & 3.9 & 3.5 & 2.3 & 9.0 & & 5.2 & 6.1 & 7.6 & \textbf{72.5} \tabularnewline
    X-vector (male)      & 1.7 & 6.5 & 1.0 & 2.9 & & 1.2 & 7.3 & 0.7 & \textbf{40.2} \tabularnewline
    \bottomrule
\end{tabular}


%% file: bios/marras_bio.tex
\begin{IEEEbiography}[{\includegraphics[width=1in,height=1.25in,clip,keepaspectratio]{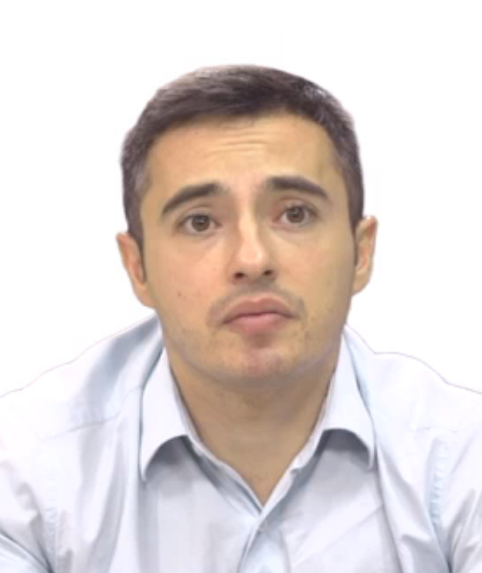}}]{Mirko Marras} is Assistant Professor at the Department of Mathematics and Computer Science of the University of Cagliari (Italy). He received his PhD Degree in Computer Science in 2020 and his MSc Degree in Computer Science (summa cum laude, 18 months) in 2016 from University of Cagliari.
Between 2020 and 2021, he spent 12 months as a Postdoctoral Researcher at the Machine Learning for Education Laboratory of EPFL (Switzerland).
His research interests focus on responsible data mining and machine learning techniques for user profiling and personalization, applied to the fields of behavioral analysis, education, business, entertainment, and social computing, with a focus on user impact. 
He is a member of (inter)national associations, including IEEE.
\end{IEEEbiography}

%% file: bios/korus_bio.tex
\begin{IEEEbiography}[{\includegraphics[width=1in,height=1.25in,clip,keepaspectratio]{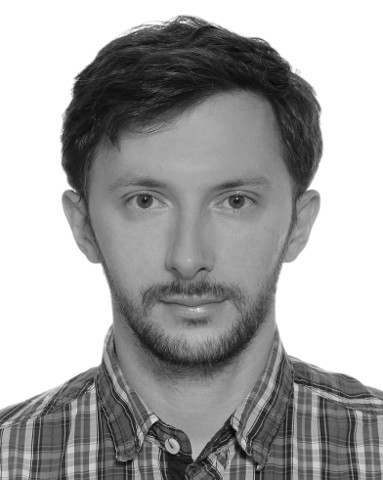}}]{Paweł Korus} received M.Sc. and Ph.D. degrees in telecommunications (with honors) from AGH University of Science and Technology, Kraków, Poland. He did his postdoctoral research at the College of Information Engineering, Shenzhen University, China. He held appointments as an assistant professor with the Department of Telecommunications, AGH University of Science and Technology, and as a research assistant professor with the Center for Cyber-Security, New York University, USA. He is currently an applied scientist with Amazon.
    
His research interests include multimedia signal processing, low-level vision and security. In 2015 he received a scholarship for outstanding young scientists from the Polish Ministry of Science and Higher Education. 
\end{IEEEbiography}

%% file: bios/jain_bio.tex
\begin{IEEEbiography}[{\includegraphics[width=1in,height=1.25in,clip,keepaspectratio]{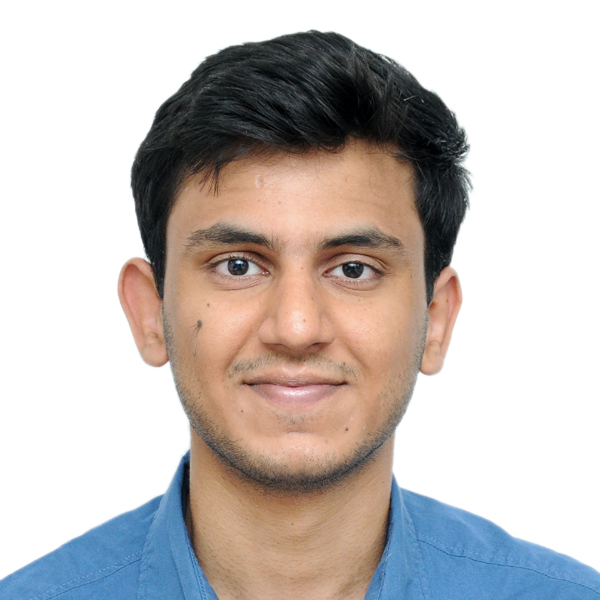}}]{Anubhav Jain} received his B.Tech. degree in Electronics and Communications Engineering from Indraprastha Institute of Information Technology, Delhi, India. He was research intern at the Idiap Research Institute, Switzerland. He is currently a PhD student at New York University, Tandon School of Engineering in the Department of Computer Science and Engineering.
His research interests include digital forensics, biometrics, and security. 
\end{IEEEbiography}

%% file: bios/memon_bio.tex
\begin{IEEEbiography}[{\includegraphics[width=1in,height=1.25in,clip,keepaspectratio]{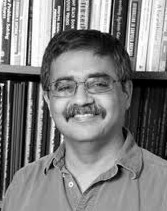}}]{Nasir Memon} is a professor in the Department of Computer Science and Engineering at NYU Tandon School of Engineering and one of the co-founders of the Center for Cyber-Security. His research interests include digital forensics, biometrics, data compression, network security and human behavior. Memon earned a Bachelor of Engineering in Chemical Engineering and a Master of Science in Mathematics from Birla Institute of Technology and Science (BITS) in Pilani, India. He received a PhD in Computer Science from the University of Nebraska.

Professor Memon has published over 250 articles in journals and conference proceedings and holds a dozen patents in image compression and security. He has won several awards including the Jacobs Excellence in Education award and several best paper awards. He has been on the editorial boards of several journals and was the Editor-In-Chief of Transactions on Information Security and Forensics. He is a Fellow of IEEE and SPIE.
\end{IEEEbiography}

%% file: tables/encoders.tex
\begin{tabular}{llrrrrrrrrrrrrrrrrrrrr}
\toprule
& \multirow{2}{*}{$\mathcal{A}~(k)$} & \multicolumn{4}{c}{AUC}                      & \multicolumn{4}{c}{EER}                         & \multicolumn{4}{c}{EER Threshold}               &             & \multicolumn{3}{c}{FRR @ FAR1\%}  &             & \multicolumn{3}{c}{FAR Threshold} \\ 
\cmidrule(lr){3-6} \cmidrule(lr){7-10} \cmidrule(lr){11-14} \cmidrule(lr){15-18} \cmidrule(lr){19-22}   
& & Raw$^1$ & Raw$^2$  & Any-10 & Avg-10 & Raw$^1$ & Raw$^2$ & Any-10 & Avg-10 & Raw$^1$ & Raw$^2$ & Any-10 & Avg-10 & Raw$^1$ & Raw$^2$ & Any-10 & Avg-10 & Raw$^1$ & Raw$^2$ & Any-10 & Avg-10 \\
\midrule
VggVox & S ($256$)           & 0.95       & 0.98           & 0.90   & 0.93   & 11.81       & 6.87            & 14.47  & 11.28  & 0.717       & 0.768           & 0.716  & 0.775  & 52.12       & 26.99           & 43.21  & 23.25  & 0.806       & 0.834           & 0.824  & 0.859  \\
ResNet 50 & S ($256$)         & 0.96       & 0.98           & 0.92   & 0.94   & 9.96        & 5.21            & 13.98  & 10.79  & 0.739       & 0.774           & 0.723  & 0.773  & 43.72       & 19.92           & 37.61  & 18.61  & 0.821       & 0.834           & 0.824  & 0.852  \\
Thin ResNet & S ($256$)      & 0.97       & 0.98           & 0.92   & 0.94   &      9.11       & 5.56            & 14.75  & 11.28  & 0.738       & 0.769           & 0.715  & 0.775  &    37.33         &         18.47        & 39.26  & 20.28  & 0.802       & 0.815           & 0.807  & 0.844  \\
XVector & F ($24$)          &     0.96        & 0.97            & 0.91   & 0.93   &       10.88      & 8.24            & 16.01  & 12.54  &    0.807         &   0.842              & 0.806  & 0.842  &       40.19      &       28.25          & 44.78  & 29.21  &        0.854     &       0.881          & 0.868  & 0.891  \\ \bottomrule
\multicolumn{20}{l}{$^0$ acoustic representations (of size $k$): (S) spectrogram; (F) filter banks; \hspace{0.1in} $^1$ tested 2.58 seconds; \hspace{0.1in} $^2$ tested on full-length } \tabularnewline
\multicolumn{10}{l}{}
\end{tabular}